\documentclass[12pt]{article}
\usepackage{epsfig}
\usepackage{a4}
\usepackage{latexsym}
\usepackage{cite}
\usepackage{amssymb}
\usepackage{graphics}

\usepackage{pslatex}
\usepackage[latin1]{inputenc}
\usepackage[T1]{fontenc}

\newcommand{\eqMel}{\raisebox{-0.07cm}{$\stackrel{\rm M}{=} $} }

\newcommand{\hspn}{{\hspace{-5mm}}}

\newcommand{\hspp}{{\hspace{3mm}}}
\newcommand{\beq}{\begin{equation}}
\newcommand{\eeq}{\end{equation}}
\newcommand{\bea}{\begin{eqnarray}}
\newcommand{\eea}{\end{eqnarray}}
\newcommand{\nn}{\nonumber}

\newcommand{\MSb}{$\overline{\mbox{MS}}$}
\newcommand{\as}{\alpha_{\rm s}}
\newcommand{\ar}{a_{\rm s}}
\newcommand{\art}{\tilde{a}_{\rm s}}

\newcommand{\ra}{\rightarrow}
\newcommand{\ep}{\epsilon}
\newcommand{\eps}{\epsilon^{\:\!2}}
\newcommand{\ept}{\epsilon^{\:\!3}}
\newcommand{\epf}{\epsilon^{\:\!4}}

\newcommand{\Ntil}{\,\widetilde{\!N}}
\newcommand{\GE}{\gamma_{\rm e}}

\begin{document}

\setlength{\parskip}{0.3cm}
\setlength{\baselineskip}{0.535cm}

\def\Ftwo{{F_{\, 2}}}
\def\FL{{F_{\:\! L}}}
\def\F3{{F_{\:\! 3}}}
\def\Qs{{Q^{\, 2}}}
\def\GeV2{{\mbox{GeV}^{\:\!2}}}
\def\x1{{(1 \! - \! x)}}
\def\z#1{{\zeta_{\:\! #1}}}
\def\ca{{C_A}}
\def\cas{{C^{\: 2}_A}}
\def\cat{{C^{\: 3}_A}}
\def\caaf{{C^{\: 4}_A}}
\def\canm{{C^{\: n-1}_A}}
\def\canmm{{C^{\: n-2}_A}}
\def\canmmm{{C^{\: n-3}_A}}
\def\canlm{{C^{\: n-\ell-1}_A}}
\def\canlmm{{C^{\: n-\ell-2}_A}}
\def\canlmmm{{C^{\: n-\ell-3}_A}}
\def\cf{{C_F}}
\def\cfs{{C^{\: 2}_F}}
\def\cft{{C^{\: 3}_F}}
\def\cff{{C^{\: 4}_F}}
\def\cffi{{C^{\: 5}_F}}
\def\cfl{{C^{\: \ell}_F}}
\def\cfn{{C^{\: n}_F}}
\def\cfnm{{C^{\: n-1}_F}}
\def\cfnmm{{C^{\: n-2}_F}}
\def\nf{{n^{}_{\! f}}}
\def\nfs{{n^{\,2}_{\! f}}}
\def\nft{{n^{\,3}_{\! f}}}
\def\caf{{C_{AF}}}
\def\cafs{{C_{AF}^{\: 2}}}
\def\caft{{C_{AF}^{\: 3}}}
\def\caff{{C_{AF}^{\: 4}}}
\def\cafn{{C_{AF}^{\: n}}}
\def\cafnm{{C_{AF}^{\: n-1}}}
\def\cafnmm{{C_{AF}^{\: n-2}}}
\def\dabc2{{d^{\:\!abc}d_{abc}}}
\def\dabcnc{{{d^{\:\!abc}d_{abc}}\over{n_c}}}
\def\dabcna{{{d^{\:\!abc}d_{abc}}\over{n_a}}}
\def\fl11{fl_{11}}
\def\flg11{fl^g_{11}}
\def\fl02{fl_{02}}
\def\b#1{{{\beta}_{#1}}}
\def\bb#1#2{{{\beta}_{#1}^{\,#2}}}

\def\B#1{{{\cal B}_{\:\!#1}}}
\def\Bb#1{{\overline{\cal B}_{\:\!#1}}}

\def\expA{{{\cal E}_A}}
\def\expF{{{\cal E}_F}}

%
\begin{titlepage}
\noindent
DESY 15-228 \hfill November 2015\\
ZU-TH 42/15 \\
LTH 1066 \\
\vspace{1.2cm}
\begin{center}
\Large
{\bf Generalized threshold resummation in inclusive DIS  \\[1mm]
  and semi-inclusive electron-positron annihilation}\\
\vspace{2cm}
\large
A.A. Almasy$^{\, a\,\ast}$, N.A. Lo Presti$^{\, b}$ and A. Vogt$^{\, c}$\\
\vspace{1cm}
\normalsize
{\it $^a$Deutsches Elektronensynchrotron DESY \\
\vspace{0.1cm}
Platanenallee 6, D--15738 Zeuthen, Germany}\\[5mm]
\normalsize
{\it $^b$Department of Physics, University of Z\"urich,\\
\vspace{0.1cm} Winterthurerstrasse 190, CH-8057 Z\"urich, Switzerland}\\[5mm]

{\it $^c$Department of Mathematical Sciences, University of Liverpool \\
\vspace{0.1cm}
Liverpool L69 3BX, United Kingdom}\\
\vspace{2.6cm}
\large
{\bf Abstract}
\vspace{-0.2cm}
\end{center}
\vspace{-0.2cm}
We present analytic all-order results for the highest three threshold
logarithms of the space-like and time-like off-diagonal splitting functions and
the corresponding coefficient functions for inclusive deep-inelastic scattering
(DIS) and semi-inclusive $e^+ e^-$ annihilation. 
All these results, obtained through an order-by-order analysis of the structure
of the corresponding unfactorized quantities in dimensional regularization, can
be expressed in terms of the Bernoulli functions introduced by one of us and 
leading-logarithmic soft-gluon exponentials. The resulting numerical 
corrections are small for the splitting functions but large for the coefficient
functions. In both cases more terms in the threshold expansion need to be 
determined in order to arrive at quantitatively reliable results.

\vspace{5mm}

\noindent
$^\ast$ {\small Address until 31 August 2013}
\end{titlepage}

%
\section{Introduction}
\label{sec:intro}
Inclusive deep-inelastic lepton-hadron scattering (DIS) and semi-inclusive
electron-positron annihilation (SIA) are phenomenologically and theoretically
important benchmark processes in perturbative quantum chromodynamics (QCD).
Data on their cross sections, respectively expressed in terms of structure
functions and fragmentation functions, form a primary source of information 
on the parton distributions and fragmentation distributions of initial- and 
final-state hadrons \cite{PDG2014}. 
The calculation of these functions in dimensional regularization is a 
standard way to determine the splitting functions governing the scale 
dependence (evolution) of the perturbatively incalculable but process-%
independent parton and fragmentation distributions. 
The coefficient functions (partonic cross sections) for DIS and SIA, 
together with those for inclusive lepton-pair and Higgs-boson production 
in hadron collisions, facilitate studies of the analytic structure of QCD 
corrections at and beyond the next-to-next-to-leading order (NNLO) of the 
renormalization-group improved perturbative expansion \cite
{vanNeerven:1991nn,Zijlstra:1991qc,Zijlstra:1992kj,Moch:1999eb,MVV6,MVV10,%
Rijken:1996vr,Rijken:1996ns,Rijken:1996np,MM06,Harlander:2002wh,%
Anastasiou:2002yz,Ravindran:2003um,Anastasiou:2015ema,Anzai:2015wma}, 
an accuracy that is harder to achieve for less inclusive quantities.

With the notable exception of the \MSb-scheme quark-quark and gluon-gluon 
splitting functions in the threshold limit \cite{Korchemsky:1989si,DMS05,%
Moch:2004pa,Vogt:2004mw,Mitov:2006ic,Moch:2007tx,AMV1}, 
all those perturbative quantities receive logarithmically enhanced 
higher-order corrections near kinematic limits. Depending on the observable 
and the kinematic region, these logarithms can require an all-order 
resummation in order to achieve phenomenologically reliable predictions. 
Knowledge of the endpoint behaviour of splitting or coefficient functions 
is also useful beyond such situations, e.g., if it can be combined with 
other partial information such as a finite number of Mellin moments of the 
splitting function or coefficient function under consideration, 
for a recent example see Ref.~\cite{MVV11}.

In this article we address the all-order resummation of threshold 
($\,x\!\ra\! 1$) logarithms of the form 
\bea
\label{x1logs}
  \x1^{\xi}\ln^{\,2n - n_0^{} - \ell\!} \x1 \quad {\rm with} 
  \quad \xi\:=\:0,\, 1 \quad {\rm and} \quad \ell \:=\:0,\, 1,\, 2 
\eea
for the $n^{\,\rm th\,}$-order splitting functions and coefficient functions 
occurring in DIS and SIA. For some coefficient functions the $\xi=0$ terms 
are subleading to the $\xi=-1$ logarithms which are the subject of the 
soft-gluon exponentiation (SGE) \cite{Sterman:1986aj,Catani:1989ne,%
Magnea:1990qg,Catani:1996yz,Contopanagos:1997nh} -- for the present status 
in DIS and SIA see \mbox{Refs.~\cite{Moch:2005ba,MV4}}; for other quantities 
the $\xi=0$ or even $\xi=1$ terms form the leading contributions.
Specifically, we complete the study of DIS in Ref.~\cite{Almasy:2010wn} by 
deriving analytic formulae for the next-to-next-to-leading logarithmic
(NNLL) $\ell = 2$ corrections and present the corresponding results for~SIA. 
A brief previous account of the latter can be found in 
Ref.~\cite{LoPresti:2012rg}, see also Ref.~\cite{Vogt:2012gb}.

The present resummations of DIS and SIA to NNLL accuracy are based on the 
NNLO fixed-order results, the structure of the unfactorized structure 
functions and fragmentation functions in $D$ dimensions and the constraints 
imposed by the all-order mass-factorization formula. 
It is worthwhile to note that the different phase-space structure of the 
Drell-Yan process and Higgs production prevents a direct generalization of 
this approach to these cases beyond the leading-logarithmic $\ell = 0$ 
accuracy of Ref.~\cite{Presti:2014lqa}.
The reader is referred to Refs.~\cite{MV3,MV5,SMVV,avLL2010,dFMMV} for an 
alternative approach (with identical results where both methods are applicable)
using physical evolution kernels, which is particularly suited for deriving 
all-$\xi$ results at a fixed order $n$.
For other, partly more formal research on subleading threshold logarithms 
see Refs.~\cite{Laenen:2008ux,Laenen:2008gt,Gardi:2010rn,Laenen:2010uz,%
White:2014qia,Bonocore:2014wua,Bonocore:2015esa,%
Grunberg:2009yi,Grunberg:2009vs,Grunberg:2011gx}.

%
\section{Threshold limits, mass factorization and resummation}
\label{sec:factorization}
\setcounter{equation}{0}
The calculations in massless perturbative QCD often take a more compact and 
transparent form in Mellin-$N$ space, defined through the integral transforms 
\beq
\label{Mtrf}
  f(N) \;=\; \int_0^1 \! dx \; x^{\,N-1}\, f(x) \quad \mbox{ and } \quad
  f(N) \;=\; \int_0^1 \! dx \left(\, x^{\,N-1} - 1 \right) f(x)_{+} \; ,
\eeq
respectively, of integrable functions and of plus-distributions.
A main advantage of working in \mbox{$N$-space} is that the ubiquitous 
(multiple) Mellin convolutions are reduced to simple product, e.g.,
\beq
\label{Mconv}
  [ f \otimes g ](x) 
  \;=\;  \int_x^1 \! \frac{dy}{y} \; f(y)\: g\!\left(\frac{x}{y}\right)
  \;\;\eqMel\;\; f(N) \, g(N) \; ,
\eeq
where $\:\eqMel\:$ indicates that the r.h.s.~is the Mellin transform of the 
previous expression. The threshold limit $x\!\ra\!1$ for the scaling variable,
e.g., Bjorken-$x$ in DIS, corresponds to the limit $N\!\ra\!\infty\,$. 
To~NNLL accuracy the dictionary between the $x$-space and $N$-space threshold 
logarithms reads
\bea
\label{LTrf}
  \Bigg( {\ln^{\,k-1\!}\x1 \over 1-x} \Bigg)_{\!+} \!
  &\!\eqMel& {(-1)^{k} \over k}\, \bigg( [S_{1-}(N)]^{\,k}
     \:+\: {1 \over 2}\, k (k-1) \z2\, [S_{1-}(N)]^{\,k-2}
     \:+\: O \big( [S_{1-}(N)]^{\,k-3} \, \big) \!\bigg) \; , \nn \\
  \ln^{\,k\!}\x1 \hspp
  &\!\eqMel& {(-1)^{k} \over N}\, \bigg( \ln^{\,k} \Ntil
    \:+\: {1 \over 2}\, k (k-1) \zeta_{\,2}\, \ln^{\,k-2} \Ntil
    \:+\: O \big( \ln^{\,k-3} \Ntil \, \big) \!\bigg) \; , \\[1mm]
  \x1 \ln^{\,k\!}\x1 \;
  &\!\eqMel& {(-1)^{k} \over N^{\,2}}\, \bigg( \ln^{\,k} \Ntil
    \:-\: k\,\ln^{\,k-1} \Ntil
    \:+\: {1 \over 2}\, k (k-1) \zeta_{\,2}\, \ln^{\,k-2} \Ntil \nn
    \:+\: O \big( \ln^{\,k-3} \Ntil \, \big) \!\bigg) \; ,
\eea
where $ S_{1-}(N) = \ln \Ntil - 1/(2N) + O(1/N^{\,2}) $ and
$\:\Ntil \,=\, N e^{\,\GE}$ with Euler's constant $\,\GE \simeq 0.577216$.
Terms suppressed by an extra power of $1/N$ have been included only in 
the first line of Eq.~(\ref{LTrf}).

In order to obtain resummed expressions for the splitting and coefficient 
functions, we address the unfactorized partonic structure functions and
fragmentation functions $T_{a,k}$ for $\,a\:=\: 2\,(T),\,L,\, \phi\,$ and 
$\,k \:=\: {\rm q,\, g}$, where $\phi$ denotes a scalar directly coupling 
only to gluons, such as the Higgs boson in the limit of a heavy top-quark 
and negligible other quark masses. These functions depend on $N$, the 
strong coupling $\as$ and, in dimensional regularization, $D = 4 - 2\ep$. 
In contrast to the physical structure functions $F_{\,2,\,L,\,\phi}$ and 
the transverse, longitudinal and $\phi$-exchange fragmentation functions 
$F_{\,T}$ and $F_{L,\,\phi}^{\,T}$, the $T_{a,k}$  are parton-level 
quantities that contain poles in the dimensional regulator $\ep$ and that 
have not been convoluted with the parton or fragmentation distributions.

Dropping all the functional dependences for brevity, these quantities can 
be factorized as
\beq
\label{Tdec}
  T_{a, k} \;\;=\;\; \widetilde{C}_{a, i} \: Z^{\,(T)}_{\,ik}
\:\: ,
\eeq
where the universal transition functions $Z^{\,(T)}_{\,ik}$ collect all 
negative powers of $\ep$ and, in the flavour-singlet DIS (`space-like') 
case, satisfy the equation
\beq
\label{PofZ}
  - \,\gamma \;\equiv\; P \;=\; \frac{d\:\! Z }{d\ln \Qs }\; Z^{\:-1} \;
  \quad {\rm with} \quad
  P\,=\,
\left(
\begin{array}{cc}
  P_{qq} & P_{qg} \\[2mm]
  P_{gq} & P_{gg} 
\end{array}
\right)\;\;,
\eeq
where $P_{ik}$ are the initial state (space-like) splitting functions; 
the $D$-dimensional coefficient functions $\widetilde{C}_{a, i}$ are 
addressed below in Eq.~(\ref{Ctilde}).
The final-state fragmentation (`time-like') transition functions 
$Z^{\,T}_{\,ik}$ satisfy an analogous equation with $\,P_{ik}$ replaced by 
$P_{ki}^{\,T}\,$. 
In Eq.~(\ref{PofZ}) we have identified, without loss of information, all 
scales with the physical scale $\Qs$ provided by the (space-like or 
time-like) momentum $q$ of the exchanged gauge boson or scalar, 
$\Qs = - q^{\,2}$ in DIS and $\Qs = q^{\,2}$ in SIA. 
This identification will by used throughout this article for both the 
renormalization and the mass-factorization scale in the \MSb\ scheme.

The (space-like and time-like) splitting functions $P_{ik}$ -- for the 
rest of this section we suppress the superscripts $(T)$ and $T$ -- can be 
expanded in powers of the strong coupling constant $\as$ as
\beq
\label{Pexp}
  P_{\:\! ik}(N,\as) \; = \;
   \sum_{n\,=\,0}^{\infty}\,\ar^{\,n+1}\, P_{\:\! ik}^{\,(n)}(N)
   \quad {\rm with} \quad
   \ar \;\equiv\; \frac{\as}{4\:\!\pi} \:\; .
\eeq
These functions are (in the time-like case: almost, with an uncertainty 
irrelevant to the present considerations) completely known to NNLO, here 
the third order in $\as$ 
\cite{Moch:2004pa,Vogt:2004mw,Mitov:2006ic,Moch:2007tx,AMV1}.
The diagonal (quark-quark and gluon-gluon) splitting functions have a 
stable form in in the large-$N$ limit \cite{Korchemsky:1989si,DMS05}, 
\beq
\label{Pii-xto1}
  P_{\:\! kk}^{\,(n-1)}(N) \; = \; \mbox{}
     -    A_{\:\! k}^{\,(n)}\, \ln\, \Ntil
  \: + \: B_{\:\! k}^{\,(n)}
  \: \pm \: C_{\:\! k}^{\,(n)}\, N^{\,-1} \ln\, \Ntil
  \: + \: {\cal O} \left( N^{\,-1} \right)
\; ,
\eeq
with the $n$-loop quark and gluon cusp anomalous dimensions related by 
$A_{\,\rm g}^{(n)}/A_{\,\rm q}^{(n)} \,=\, \ca/\cf$ at $n \leq 3$, a fact 
often referred to as {Casimir scaling}.
The coefficient of $N^{\,-1} \ln\, \Ntil$ can be expressed in terms of 
lower-order cusp anomalous dimensions \cite{DMS05}, and differs in sign 
between the DIS and SIA cases.
The large-$N$ behaviour of the off-diagonal (quark-gluon and gluon-quark) 
splitting functions, on the other hand, is characterized by a double-%
logarithmic higher-order enhancement,
\beq
\label{Pij-xto1}
  P_{\:\! i\neq k}^{\,(n)}(x) \; = \; \frac{1}{N}\:
   \sum_{\ell\,=\,0}^{2n}\, D_{\:\! ik}^{\,(n,\ell)}\,\ln^{\,2n-\ell\,} \Ntil
   \: + \: {\cal O}\left( \frac{1}{N^{\,2}} \ln^{\,m\,} \Ntil \right) \; ,
\eeq
where the terms with $\,\ell=0\,$ form the leading-logarithmic (LL) 
approximation \cite{AV2010}, those with $\,\ell=1\,$ the 
next-to-leading-logarithmic (NLL) contributions etc. 
Analytic results for the $\ell \leq 2$ coefficients in Eq.~(\ref{Pij-xto1})
will be presented in Sections 4 and 5 below.

After expanding the transition functions in powers of the strong coupling as
\beq
\label{Zexp}
  Z_{ik} \;=\; \sum_{n\,=\,1}^{\infty} \ar^{\,n}\,Z_{ik}^{\,(n)}
\eeq
Eq.~(\ref{PofZ}) can be solved order by order in $\ar$. 
With $\,\gamma_{\,n}^{} = - P^{\,(n)}$ this results in
\bea
\label{Z4ofP}
  \; Z \!\!  &\!=\!& \!\! 1 
  \:+\: \ar \:{1 \over \ep}\: \gamma_0^{}
  \:+\: \ar^{\,2} \bigg\{ \, {1 \over 2\:\!\eps}\,
        \left( \gamma_0^{} - \beta_0 \right) \gamma_0^{}
        \:+\: {1 \over 2\:\!\ep}\: \gamma_1^{} \bigg\}
  \nn \\[0.5mm] & & \mbox{\hspn}
  \:+\:  \ar^{\,3} \bigg\{ \, {1 \over 6\:\!\ept}
        \left( \gamma_0^{} - \beta_0 \right)
        \left( \gamma_0^{} - 2\:\!\beta_0 \right) \gamma_0^{}
        \:+\: {1 \over 6\:\!\eps}
        \Big[ \left( \gamma_0^{} - 2\:\!\beta_0 \right)
        \gamma_1^{} + \left( \gamma_1^{} - \beta_1 \right)
        2\:\!\gamma_0^{} \Big] + {1 \over 3\:\!\ep}\: \gamma_2^{}
        \bigg\}
  \nn \\[0.5mm] & & \mbox{\hspn}
  \:+\:  \ar^{\,4} \bigg\{ \, {1 \over 24\:\!\epf}
        \left( \gamma_0^{} - \beta_0 \right)
        \left( \gamma_0^{} - 2\:\!\beta_0 \right)
        \left( \gamma_0^{} - 3\:\!\beta_0 \right) \gamma_0^{}
  \nn \\[0.5mm] & & \mbox{\hspp}
        +\: {1 \over 24\:\!\ept} \Big[
          \left( \gamma_0^{} - 2\:\!\beta_0 \right)
          \left( \gamma_0^{} - 3\:\!\beta_0 \right) \gamma_1^{}
        + \left( \gamma_0^{} - 3\:\!\beta_0 \right)
          \left( \gamma_1^{} - \beta_1 \right) 2 \gamma_0^{}
        + \left( \gamma_1^{} - 2\:\!\beta_1 \right)
          \left( \gamma_0^{} - \beta_0 \right) 3 \gamma_0^{}
        \Big]
 \quad \nn \\[0.5mm] & & \mbox{\hspp}
        +\: {1 \over 24\:\!\eps} \Big[
          \left( \gamma_0^{} - 3\:\!\beta_0 \right) 2 \gamma_2^{}
        + \left( \gamma_1^{} - 2\:\!\beta_1 \right) 3 \gamma_1^{}
        + \left( \gamma_2^{} - \beta_2 \right) 6 \gamma_0^{}
        \Big]
        \:+\: {1 \over 4\:\!\ep}\: \gamma_3^{}
        \bigg\}
  \:\;+\;\; \ldots \;\; .
\eea
Here $\beta_n$ are the usual N$^{n}$LO coefficients of the beta function of 
QCD, with $\beta_0 \:=\: 11/3\,\ca - 2/3\: \nf\,$, where $\nf$ is the number
of effectively massless quark flavours and $\,\ca \,=\, n_{\:\!\rm colours} 
\,=\, 3$.
There is no general closed all-order form of this result; however for
the present large-$N$ limit an explicit, if still rather lengthy form has
been given in Eq.~(2.7) -- (2.13) of Ref.~\cite{Almasy:2010wn}.
Schematically, the leading-logarithmic behaviour of $Z$ in $N$-space  
corresponding to Eqs.~(\ref{Pii-xto1}) and (\ref{Pij-xto1}) is given by
\bea
\label{Zlog}
  Z_{kk}^{\,(n)} &\!\sim\!& 
  (1 \,+\, \ep\ln \Ntil \,+\, \dots \,+\, \ep^{\,n-1}\ln^{\,n-1} \Ntil\, )\; 
  \ep^{\,-n}\,\ln^{\,n}\Ntil
\;\; , \nn \\[1mm]
  Z_{i\neq k}^{\,(n)} &\!\sim\!& 
  (1 \,+\, \ep\ln \Ntil \,+\, \dots \,+\, \ep^{\,n-1}\ln^{\,n-1} \Ntil\, )\;
  \ep^{\,-n}\, N^{\,-1}\, \ln^{\,n-1} \Ntil
  \;\; .
\eea

The process-dependent $D$-dimensional coefficient functions 
$\widetilde{C}_{a, i}$ in Eq.~(\ref{Tdec}) include contributions with all 
non-negative powers of $\ep$. Their expansion in powers of $\as$ and $\ep$
can be written~as
\beq
\label{Ctilde}
  \widetilde{C}_{a, i} \;=\; 
        \delta_{\,a\:\!\gamma}\, \delta_{\,i\:\!\rm q} 
  \:+\: \delta_{\,a\:\!\phi}\, \delta_{\,i\:\!\rm g}
  \:+\: \sum_{n\,=\,1}^{\infty}\, \ar^{\,n}\, \sum_{k\,=\,0}^{\infty}\: 
        \ep^{\,k} c_{a, i}^{\,(n,k)} \;. 
\eeq
Here the index $\gamma$ of the Kronecker-delta indicates that 
$\,\delta_{\,a\:\!\gamma}\,$ is equal to one if $a=2$ in DIS and $a=T$ in 
SIA, and zero for $a=\phi$ or $L$.
The $\ep$-independent contributions, $c_{a, i}^{\,(n,0)} \:\equiv\; 
c_{a, i}^{\,(n)}$, are the $n^{\,\rm th\,}$-order coefficient functions 
entering the physical structure functions and fragmentation functions. 

The quark coefficient functions for the gauge-boson-exchange structure and 
fragmentation functions $F_{\,2,\,T}$ and the gluon coefficient function 
for the scalar-exchange structure and fragmentation function $F_\phi$, 
also referred to as `diagonal' coefficient functions, are dominated in the 
large-$N$ limit by Mellin-transformed plus-distributions with a the 
double-logarithmic enhancement,  
\beq
\label{Cdiag-xto1}
  c_{a, k}^{\,(n)}(x) \;=\; 
  \sum_{\ell\,=\,0}^{2n}\, D_{a,\, k}^{\,(n,\ell)}\, \ln^{\,2n-\ell\,} \Ntil
  \; + \; \frac{1}{N} \sum_{\ell\,=\,0}^{2n-1}\, E_{a,\, k}^{\,(n,\ell)}\, 
          \ln^{\,2n-1-\ell\,} \Ntil
  \; + \; {\cal O}\left( \frac{1}{N^{\,2}} \ln^{\,m\,} \Ntil \right) 
\eeq
for $\{a,\, k \} \:=\: \{2, q\}, \{T,q\}$ and $\{\phi,g\}$.
The first sum in Eq.~(\ref{Cdiag-xto1}) includes the contributions that 
are resummed by the soft-gluon exponentiation \cite{Sterman:1986aj,%
Catani:1989ne,Magnea:1990qg,Catani:1996yz,Contopanagos:1997nh}, 
with the coefficients $D_{a,\, k}^{\,(n,0)}$ \dots $D_{a,\, k}^{\,(n,2n-2)}$ 
at order $n$ fixed by lower-order information.
At present the coefficients of the six highest logarithms are known 
analytically, and for the seventh only the (numerically small) 
contribution from the four-loop cusp anomalous dimension is missing; 
see Ref.~\cite{Moch:2005ba} and Ref.~\cite{MV4} for the respective
gauge-boson exchange DIS and SIA results.

Complete all-order results for the LL, NLL and NNLL ($\ell \leq 2$) 
coefficients $E_{2,\, q}^{\,(n,\ell)}$ and $E_{T,\, q}^{\,(n,\ell)}$ in 
Eq.~(\ref{Cdiag-xto1}) were derived in Ref.~\cite{MV5} from a conjecture 
on the respective physical evolution kernels. 
In the present approach, we were able to verify those results (and hence the 
underlying conjecture) and to extend them to the $\ell = 3$ coefficients, 
thus fixing the corresponding unknown coefficients in Ref.~\cite{MV5} as 
$\xi_{\rm DIS_4} = \xi_{\rm SIA_{\,4}}$ \cite{Vogt:2012gb}.
This agrees with the result obtained in Ref.~\cite{Grunberg:2009vs}.

The remaining coefficient functions for the observables in 
Eq.~(\ref{Cdiag-xto1}) start only at order $\as$ and can be considered as
`off-diagonal' quantities (the off-diagonal splitting functions arise from
their unfactorized counterparts). Their leading large-$N$ behaviour is 
completely analogous to the subleading $1/N$ contributions in 
Eq.~(\ref{Cdiag-xto1}), viz
\beq
\label{Cak-xto1}
  c_{a, k}^{\,(n)}(x) \;=\; \frac{1}{N} \sum_{\ell\,=\,0}^{2n-1}\, 
        D_{a,\, k}^{\,(n,\ell)}\, \ln^{\,2n-1-\ell\,} \Ntil
  \; + \; {\cal O}\left( \frac{1}{N^{\,2}} \ln^{\,m\,} \Ntil \right)
\eeq
for $\{a,\, k \} \:=\: \{2, g\}, \{T,g\}$ and $\{\phi,q\}$.
Besides their splitting-function counterparts in Eq.~(\ref{Pij-xto1}) above,
the all-order determination of the $l \leq 2$ coefficients constitutes in 
Eq.~(\ref{Cak-xto1}) a main objective of this article; the corresponding DIS 
and SIA results can be found in Sections 4 and 5. 

Finally the large-$N$ expansion of the $n^{\,\rm th\,}$-order coefficient 
functions for the structure function $F_L$ and the fragmentation function 
$F_L^{\,T}$ have the form, for $k \,=\,q,\,g$,
\beq
\label{CLk-xto1}
  c_{L, k}^{\,(n)}(x) \;=\; 
    \frac{1}{N^{\,1+\,\delta_{\:\!k\rm g}}} \sum_{\ell\,=\,0}^{2n-2}\,
    D_{L,\, k}^{\,(n,\ell)}\, \ln^{\,2n-2-\ell\,} \Ntil
  \; + \; {\cal O}\left( \frac{1}{N^{\,2+\,\delta_{\:\!k\rm g}}} 
    \ln^{\,m\,} \Ntil \right) \; .
\eeq
In the quark cases our present calculations verify the physical-kernel 
based results of Refs.~\cite{MV3,MV5} for the coefficients up to NNLL 
accuracy ($l \leq 2$), but cannot add to those results. The closed form of 
the NNLL resummation for the case of DIS has been given already in Eq.~(6.3) 
of Ref.~\cite{Almasy:2010wn}; the~corresponding SIA result will be presented 
in Section 5.

The $k>0$ coefficients $c_{a, i}^{\,(n,k)}$ in the $\ep$-expansion of the 
$D$-dimensional coefficient function in Eq.~(\ref{Ctilde}) are enhanced by 
factors $\ln^{\,k} \Ntil$ with respect to the four-dimensional coefficient 
functions discussed in Eqs.~(\ref{Cdiag-xto1}) -- (\ref{CLk-xto1}), i.e.,
the pattern for the non-negative powers of $\ep$ is the same as in 
Eq.~(\ref{Zlog}) for the $1/\ep$ poles. Consequently the unfactorized
structure functions and fragmentation functions $T_{a,k}$ in Eq.~(\ref{Tdec})
exhibit the same structure over all powers of $\ep$.

Disregarding the logarithms and (except for $T_{L,g}$) terms
$\widetilde{C}_{a, i} \: Z_{\,ik}$ that contribute 
only at order $1/N^{\,2}$, the large-$N$ behaviour of the unfactorized
structure functions and fragmentation functions can be summarized as 
follows, with $\gamma = 2$ in DIS and $\gamma = T$ in SIA,
\bea
 T_{\,\gamma,q}^{} \;\simeq\; \widetilde{C}_{\,\gamma, q}\,Z_{qq} 
             \;\sim\; \mathcal{O}(1)
\quad\; &\;\rightarrow\;&
 T_{\,\gamma,g}^{} \;=\; \widetilde{C}_{\,\gamma, q}\,Z_{qg} 
             \,+\,\widetilde{C}_{\,\gamma, g} \,Z_{gg} 
      \;\;\sim\;\;\mathcal{O}(1/N)
\;\;,\nn\\[1mm]
 T_{\phi,g} \;\simeq\; \widetilde{C}_{\phi,g}\,Z_{gg} \;\sim\; \mathcal{O}(1)
\quad\; &\;\rightarrow\;&
 T_{\phi,q} \;=\; \widetilde{C}_{\phi,g}\,Z_{gq}
         \,+\,\widetilde{C}_{\phi,q}\,Z_{qq} 
      \;\;\sim\;\;\mathcal{O}(1/N) 
\;\;,\label{eq:T-log}\\[1mm]
  T_{L,q}^{} \;\simeq\; \widetilde{C}_{L,q}\,Z_{qq} \;\sim\; \mathcal{O}(1/N)
 &\;\rightarrow\;&
  T_{L,g} \;=\; \widetilde{C}_{L,q}\,Z_{qg} 
          \,+\, \widetilde{C}_{L,g}\,Z_{gg} 
      \;\;\sim\;\;\mathcal{O}(1/N^{\:\!2})
\;\;.\nn
\eea
Here the arrows indicate that the resummation of the quantities on the 
extract the all-order `off-diagonal' coefficient functions. 

Once the unfactorized structure function is known at order $\ar^{\,n}$, 
it is possible to extract the coefficient function $c_{a, i}^{\,(n,0)}$,
provided that the lower-order contributions $c_{a, i}^{\,(m,k)}$ are known
to a sufficiently high power of $\ep$.
In particular, the calculation of $T_a$ to order $\ar^{\,\ell \leq n}$ and 
$\ep^{\,n-\ell}$ is required for the extraction of the coefficient function 
at order $\ar^{\,n}$.  
On the other hand, one can see from Eq.~(\ref{Z4ofP}) that a full 
N$^{\,n-1}$LO result completely fixes the highest $n$ powers of $1/\ep$ to 
all orders in $\ar$.
In order to be able to extract the splitting and coefficient functions from 
the unfactorized structure function and fragmentation functions at all orders 
(at the logarithmic accuracy under consideration), it is thus necessary to 
consider the $D$-dimensional coefficient functions at all powers of $\ep$. 
 
It was noted in Ref.~\cite{Almasy:2010wn} that the $\ar^{\,n}$ 
contributions to the unfactorized structure functions $T_{2,\,g}^{\,}\,$, 
$T_{\phi,\,q}^{\,}\,$ and $T_{L,\,k}^{\,}\,$ in Mellin space can be written as
\beq
\label{T2pABC}
  T^{\,(n)}_{a,k}(N) \;=\; 
  \frac{1}{N^{\,1+\delta_{aL}\delta_{kg}}\:\ep^{\,2n-1-\delta_{aL}}}\; 
  \sum_{i=0}^{n-1}\, 
  \left( 
         A^{\,(n,i)}_{a,k} \,+\, 
  \ep\,  B^{\,(n,i)}_{a,k} \,+\,  
  \eps\, C^{\,(n,i)}_{a,k} \,+\, \ldots\,
  \right)\, \exp \left(\ep\, (n-i)\, \ln N \right) \; .
\eeq
The fact that the right-hand-side features double poles in $\ep$, whereas the 
mass-factorization \mbox{formula} ensures that only single poles appear in 
the unfactorized structure function, imposes constraints on the coefficients 
$\,A^{\,(n,i)}_{a,k}\,$, $\,B^{\,(n,i)}_{a,k}\,$ and  $\,C^{\,(n,i)}_{a,k}\,$.
In the large-$N$ limit, once the exponential on the r.h.s.~is expanded, 
only the leading logarithmic (LL) coefficients $A^{\,(n,i)}_{a,k}$ 
appear in all the vanishing $\,\epsilon^{-2n+1}\,,\dots,\,\epsilon^{-n-1}$ 
terms, so we immediately have $\,n-1\,$ relations for these coefficients.
The~next-to-leading logarithmic (NLL) coefficients $B^{\,(n,i)}_{a,k}$ will 
appear in all double-pole terms except the one proportional to  
$\,\epsilon^{-2n+1}$, resulting in $\,n-2\,$ equations;
at next-to-next-to-leading logarithmic (NNLL) level there are $\,n-3\,$ 
relations for the coefficients $C^{\,(n,i)}_{a,k}$ and so on.
In general, the cancellation of the double poles in $\epsilon$ provides 
$\,n-1-m\,$ relations between the $n$ N$^{\:\!m\:\!}$LL coefficients. 
The first $k+1$ powers of $\ep^{-1}$ with non-vanishing coefficients are 
fixed by a full N$^{\:\!k\:\!}$LO calculation, as discussed at the end of 
the previous section, leading to a total of $\,n-m+k\,$ relations. 
Since at order $\ar^{\,n}$ each linear system has $n$ unknowns (notice the 
sum over the index $i$ in Eq.~(\ref{T2pABC})), the coefficients up to the 
N$^{\:\!k\:\!}$LL terms are fixed in terms of the N$^{\:\!k\:\!}$LO results.
The  N$^{\:\!m\:\!}$LL coefficients with $m<k$  are over-constrained to 
all orders, providing a check on the correctness of Eq.~(\ref{T2pABC}). 
The same holds for $m=k$ beyond oder $k$.

Having determined the coefficients of Eq.~(\ref{T2pABC}), the 
mass-factorization formula (\ref{Tdec}), together with the all-order 
solution of Eq.~(\ref{PofZ}) in the large-$N$ limit, see Eqs.~(2.7) -- 
(2.13) of Ref.~\cite{Almasy:2010wn}, \mbox{allows} the iterative 
determination of the coefficients $D_{\:\! ik}^{\,(n,\ell \,\leq\, 2)}$ 
in Eq.~(\ref{Pij-xto1}) and $D_{a,\, k}^{\,(n,\ell \,\leq\, 2)}$ in 
(\ref{Cak-xto1}) and (\ref{CLk-xto1}) to, in principle, any order in $\as$. 
It may be worthwhile to note that the all-order expressions for 
$Z_{ik}^{\,(n)}$ are not a superfluous luxury: the results of the mass 
factorization are required to a very high order, beyond what can be easily 
achieved by an order-by-order brute-force solution of (\ref{PofZ}), for the 
reconstruction of the NNLL analytic forms presented in the next two sections.

%
\section{The NNLL corrections in DIS in closed form}
\label{DISresults}
\setcounter{equation}{0}
In Ref.~\cite{AV2010} it was found that the LL contributions to the 
resummed off-diagonal splitting and coefficient functions can be expressed
in a closed form in terms of an apparently new function $\B{0}$, 
\bea
\label{B0def}
  \B{0}(x) \; = \;
  \sum_{n\,=\,0}^\infty \;\frac{B_n}{(n!)^2}\; x^{\,n} 
  &\!= \!& 
  1 \,-\: {x \over  2}
  \; - \;\sum_{n=1}^\infty \,\frac{(-1)^n}{[(2n)!]^2} \; |B_{2n}|\, x^{\,2n}
\nn \\ &\!= \!&
  1 \:-\: {x \over  2} \:-\: 2 \sum_{n=1}^\infty \:
  {(-1)^n \over  (2n)!} \,\zeta_{2n}^{} \Big( {x \over  2\pi} \Big)^{\!2n}
\; .
\eea
Here $B_n$ are the Bernoulli numbers as normalized in Ref.~\cite{AbrSteg};
$\zeta_{n}$ denotes Riemann's \mbox{$\zeta$-function}. Due to $\zeta_{n} 
\ra 1$ for $n \ra \infty$, the Taylor series (\ref{B0def}) absolutely 
converges for all values of $x$.

This resummation was extended to the NLL and NNLO contributions to the 
splitting functions for the evolution of parton distributions and the 
coefficients functions for inclusive DIS in Ref.~\cite{Almasy:2010wn}. 
However, with the exception of the longitudinal structure function, closed
forms were found only for the NLL corrections. Besides Eq.~(\ref{B0def}),
these expressions involve the generalizations
\beq
\label{Bkdef}
  \B{k}(x) \;=\;
  \sum_{n\,=\,0}^\infty \;\frac{B_n}{n!(n+k)!}\; x^{\,n} \, ,\quad
  \B{-k}(x) \;=\;
  \sum_{n\,=\,k}^\infty \;\frac{B_n}{n!(n-k)!}\; x^{\,n} \:\: 
\eeq
which are related to $\B{0}$ by
\beq
\label{Bkdiff}
  \frac{d^{\,k}}{dx^{\,k}}\; (x^{\,k} \B{k}) \;=\; \B0
\:\: , \quad
  \frac{d^{\,k}}{dx^{\,k}}\; \B0 \;=\; {1 \over  x^{\,k}}\; \B{-k}
\:\: .
\eeq
Specifically, the NLL terms can be expressed by $\B{k}$ with 
$-2 \leq k \leq 1$; plots of these functions can be found in Fig.1 of
Ref.~\cite{Almasy:2010wn}. On the other hand, the NNLL corrections for
$P_{qg}$, $P_{qg}$, $C_{2,g}$ and $C_{\phi,q}$ were only give via tables
to order $\as^{18}$ for the off-diagonal splitting functions and to 
order $\as^{12}$ for the corresponding coefficient functions.

By extending the calculations to a considerably higher order than before,
and thus generating an over-constrained system of linear equations for a
suitably general ansatz, we have been able to derive the hitherto missing 
closed forms. They are much more complicated than their LL and NLL 
counterparts, but involve the same ingredients: the functions $\B{k}$, for 
the coefficient functions in combination with the LL exponentials for the 
soft-gluon resummation of $C_{2,q}$ and $C_{\phi,g}$.

\noindent
The large-$N$ space-like gluon-quark and quark-gluon splitting functions 
read, at NNLL accuracy,
\bea
\label{eq:pSqgNNL} 
 N\:\!P_{\rm qg}^{\,\rm S}(N,\as) &\!=\!& 
	  2\* \nf\*\:\! \ar\*\, \B0 
\nn \\[0.8mm] & & \hspace{-2.5cm}
	+ \,\nf\*\ar^2\* \ln \Ntil\:\* 
    \Big\{
    		\left( 6\* \cf - \b0 \right)\*\bigg[
    			  \B1 
    			+ \,2\art^{-1}\:\* \B{-1} 
    		  \bigg]
    		+\: \b0\art^{-1}\:\* \B{-2} 
    	\Big\}
\nn \\[0.8mm] & & \hspace{-2.5cm}
 	+\, \frac{\nf\*\ar^2}{48\*\,\caf}\* 
 	\Bigg\{
  		\, 108\* \, \cfs\* \Big[
  			  2 \* \art \* \B2
	  		- \,4 \* \B1
	  		+ 5 \* \B0
  			+ \, 2\art^{-1} \* \B{-1} 
  			+ \, 4\art^{-1} \* \B{-2}
  		  \Big] 
\nn \\[1mm] & & \hspace{-1.9cm}
  		- \,36 \*\, \b0\* \, \cf \* \Big[ 
  			  \art \* \B2 
			- \,3 \* \B1 
			+ 4 \* \B0
			-  \B{-1}
  			+ \,2\art^{-1} \* \B{-1} 
  			+ \,\art^{-1} \* \B{-2}
  			- \,2\art^{-1} \* \B{-3}
  		  \Big]
\nn \\[1mm] & & \hspace{-1.9cm}
  		+   \bb02\* \Big[
  			  2 \* \art \* \B2 
  			- 12 \* \B1 
  			+ 12 \* \B0 
  			- 6 \* \B{-1}
  			- \,12\art^{-1} \* \B{-2}
  			- \,4\art^{-1} \* \B{-3}  
			+ \,3\art^{-1} \* \B{-4}
		  \Big]
\nn \\[1mm] & & \hspace{-1.9cm}
  		+ \, 80\* \,\caf \* \b0 \* \Big[ 
  			  \art \* \B2 
  			- \, 4 \* \B1 
  			+ \, 4 \* \B0  
  			+  \B{-1}
  		  \Big]
\nn \\[1mm] & & \hspace{-1.9cm}
  		- \, 32\* \,\caf\* \,\cf \* \Big[
  			  (19-3 \* \z2) \* \art \* \B2 
  			- 34 \* \B1 
  			+ (13+6 \* \z2) \* \B0 
  			- (2-3 \* \z2) \* \B{-1}
  		  \Big]
\nn \\[1mm] & & \hspace{-1.9cm}
  		+ \, 32\* \,\cafs\*\, \Big[
  			  (2+3 \* \z2) \* \art \* \B2 
  			+ 4 \*(1+3 \* \z2) \* \B1
  			+ 2 \* (1-6 \* \z2) \* \B0
  			+ (2-3 \* \z2) \* \B{-1}
  		  \Big]
	\Bigg\}
\qquad
\eea
(this expression was already presented in Ref.~\cite{Vogt:2012gb}) and
\bea
\label{eq:pSgqNNL} \hspn
N\:\!P_{\rm gq}^{\,\rm S}(N,\as) &\!=\!& 
	  2\*\cf\*\:\!\ar\*\,\Bb0
\nn \\[0.8mm] & & \hspace{-2.5cm}
 	+ \,\cf\*\ar^2\* \ln \Ntil\:\*
    \Big\{ 
    		\left( 14\*\,\cf - 8\*\:\!\ca - \b0 \right)\*\:\Bb1
 		+ \:6\*\left( 2\*\:\!\cf - \:\!\b0 \right)\*\:\art^{-1}\*\:\Bb{-1}
  		- \: \b0\art^{-1}\*\:\Bb{-2}
  	\Big\}
\nn \\[1mm] & & \hspace{-2.5cm}
 	+ \frac{\cf\*\ar^2}{48\*\,\caf}\* 
 	\Bigg\{
	     108\* \, \cfs\* \Big[
			  2 \* \art \* \Bb2
   			- 8 \* \Bb1
   			+ 7 \* \Bb0
			+ 2\art^{-1} \* \Bb{-1}
   			+ 4\art^{-1} \* \Bb{-2}
   		  \Big] 
\nn \\[1mm] & & \hspace{-1.9cm}
		- 36\* \, \b0\* \, \cf \* \Big[  
	   		  \art \* \Bb2
   			- 19 \* \Bb1
   			+ 16 \* \Bb0
   			+  \Bb{-1}
   			+ 6\art^{-1} \* \Bb{-1}
   			+ 15\art^{-1} \* \Bb{-2}
   			+ 2\art^{-1} \* \Bb{-3}
   		  \Big]
\nn \\[1mm] & & \hspace{-1.9cm}
		+ \bb02 \* \Big[ 
	     	  2 \* \art \* \Bb2
  			- 108 \* \Bb1
			+ 84 \* \Bb0
  			+ 6 \* \Bb{-1}
  			+ \art^{-1}\*\left(
  			48\* \Bb{-1}
  			+ 156\* \Bb{-2} 
  			+ 44\* \Bb{-3}
  			+ 3\* \Bb{-4} \right)
  		  \Big]
\nn \\[1mm] & & \hspace{-1.9cm}
   		- 32 \* \caf \* \b0 \* \Big[
   			  \art \* \Bb2
   			- 7 \* \Bb1
			- 9 \* \Bb0
   			- 4 \* \Bb{-1}
   		  \Big]
\nn \\[1mm] & & \hspace{-1.9cm}
   		+ 32\* \, \caf\* \, \cf \* \Big[
   			  (10-3 \* \z2) \* \art \* \Bb2 
   			- 25 \* \Bb1
   			+ 3 \* (1-2 \* \z2) \* \Bb0
  			+ (2-3 \* \z2) \* \Bb{-1}
  		  \Big]
\nn \\[1mm] & & \hspace{-1.9cm}
  		- 32\* \, \cafs\* \,  \Big[
  			  (2-9 \* \z2) \* \art \* \Bb2 
  			- 4 \* (5+ 3 \* \z2) \* \Bb1
  			+ 6 \* (1+2\*\z2) \* \Bb0
  			- (2-3 \* \z2) \* \Bb{-1}
  		  \Big]
	\Bigg\} 
\eea
with the shorthand notations
\beq
\label{atilde}
  \art \;\equiv\; 4\:\!\ar\,\caf\,\ln^2\Ntil, 
  \quad \caf \;\equiv\; \ca - \cf \; .
\eeq
Further, we have suppressed everywhere the argument of the ${\cal B}_k$ 
functions and used $\B{k}\equiv\B{k}(\art)$ and $\Bb{k}\equiv\B{k}(-\art)$. 
The respective first lines in Eqs.~(\ref{eq:pSqgNNL}) and (\ref{eq:pSgqNNL}) 
represent the LL result \cite{AV2010}, the second lines the NLL result 
\cite{Almasy:2010wn} and the remaining parts are the new NNLL expressions.

The resummed expressions for the gluon coefficient functions for $F_{\,2}$
and the quark coefficient function for $F_{\,\phi}$ at NNLL accuracy read
\bea
N\:\!C_{2,\rm g}(N,\as) &\!=\! & 
	  \frac{\nf}{2\*\,\caf\*\ln \Ntil}\*\:
		  \left[\B0\* \expF - \expA \right]
  	+ \: \frac{\nf\*\left(\b0 - 3\*\cf\right)}{8\*\,\cafs\*\ln^2\Ntil}\* 
  		  \left[ \B0\* \expF - \expA \right]
\nn \\[1mm] & & \hspace{-2.5cm} 
 	+ \: \frac{\nf\*\ar}{4\*\,\caf}\*\: 
 	\Big\{
 		  6\*\cf\*\left[
 		  	  \left(
 		  	  	  \B1
 		  	 	+ \B0
 		  	 	+ 2\art^{-1}\*\B{-1}
 		  	 \right) \*\expF
 		- 2\*\expA
 		\right]
  		- \, 8\*\caf\*\expA 
				- \,\b0\*\big[
			  \big(
			  	  \B1
\nn \\[1mm] & & \hspace{-2cm} \mbox{}
			  	+ 4\art^{-1}\*\B{-1}
			  	- \art^{-1}\*\B{-2}
			  \big)\*\expF 
		- \expA
		\big]
  	\Big\} 
	+ \, \frac{\nf\*\ar^2\*\b0\*\ln^2\Ntil}{3\*\,\caf}\*
  		  \left[\cf\*\B0\*\expF -  \ca\*\expA \right]
		  	\nn \\[1mm] & & \hspace{-2.5cm} 
	+ \: \frac{\nf}{32\*\,\caft\* \ln^3\, \Ntil}\* 
	\Big\{ 
		 \bb02\*\bigg[ 
			  \frac{1}{3}\*\B{-3}
			+ \frac{1}{8}\* \B{-4}
		\bigg]\* \expF  
		- \b0\*(\b0 - 3\*\cf)\*\bigg[ 
   	  		  \B{-2}
   	  		+ \B{-3}
   	  	\bigg]\* \expF 
\nn \\[1mm] & & \hspace{-2cm} \mbox{}
		 + (\b0 - 3\*\cf)^2\*\bigg[
			  \left(
			  	  \B0
			  	- \B{-1} 
			  	+ 2\* \B{-2}
			  \right)\* \expF 
			- \expA 
		\bigg]
	\Big\}
\nn \\[1mm] & & \hspace{-2.5cm} 
    + \:\frac{\nf\*\ar}{96\*\,\cafs\* \ln\, \Ntil} \*
    \Big\{
		- 3\*\bb02\* \bigg[
			  \left(
			  	  4 \* \B1
			  	- 4 \* \B0
			  	+ \B{-1}
			  \right) \* \expF 
			  - 2\* \, \expA 
		  \bigg]
	\nn \\[1mm] & & \hspace{-2cm}   \mbox{}
     	- 54\* \cfs\* \bigg[
     		  \left(
     		  	  6 \* \B1
     		  	- 3 \* \B0
     		  	- 4 \* \B{-1}
     		  \right) \* \expF 
     		  - 4\* \, \expA 
     	\bigg]
     	- 40\*\b0 \* \caf\* \bigg[
     		  \bigg(
     		  	  4 \* \B1
     		  	- 2 \* \B0
	\nn \\[1mm] & & \hspace{-2cm} \mbox{}
     		  	- \B{-1}
     		  \bigg) \* \expF 
     		  - 2 \* \expA
     	\bigg]	
     	+ 18\*\b0\* \, \cf\* \bigg[
     		  \left(
     		  	  6 \* \B1
     		  	- 4 \* \B0
     		  	- 3 \* \B{-1}
     		  	+ \B{-2}
     		  \right) \* \expF 
     		  - 4\* \, \expA
     	\bigg]
\nn \\[1mm] & & \hspace{-2cm} \mbox{}
     	+ 16\*\cf \* \caf\* \bigg[
     		  \left(
     		  	  34\* \B1 
     		  	- 2\* \, (22+3 \* \z2) \* \B0
     		  	+ (2-3 \* \z2) \* \B{-1}
     		  \right) \* \expF 
    			  + 2\* \, (5+3\* \, \z2)\* \, \expA 
    		\bigg]
\nn \\[1mm] & & \hspace{-2cm} \mbox{}
     	+  16\*\cafs \* \bigg[
     		  \left(
     		  	  2\*(1+3 \* \z2)\*(2 \* \B1-\B0)
     		  	+ (2-3 \* \z2) \* \B{-1}
     		  \right) \* \expF 
    			  - 2\*(1+3 \* \z2) \* \expA 
    		\bigg] 
    	\Big\}
\nn \\[1mm] & & \hspace{-2.5cm} 
	+  \:\frac{\nf\*\ar^2 \* \ln\, \Ntil}{24\*\,\cafs}\* 
	\Big\{
		  \bb02 \* \cf\* \bigg[
			  \left(
			  	  2 \* \B0
			  	- 4 \* \B{-1}
			  	+ \B{-2}
			  \right)\* \expF 
			  - 2\* \, \expA 
		\bigg]
\nn \\[1mm] & & \hspace{-2cm} \mbox{}
		- 6\*\b0 \* \cfs\* \bigg[
			  \left(
			  	  \B0
			  	- 2 \* \B{-1}
			  \right)\*\, \expF 
			  - \expA
		\bigg]
     	+  \bb02 \* \caf\* \bigg[
     		    \B2\* \, \expF 
     		  - \expA
     	\bigg]
\nn \\[1mm] & & \hspace{-2cm} \mbox{}
     	-  2\* \b0 \* \cf \* \caf\* \bigg[
     		  \left(
     		  	  9\* \B2
     		  	+ 9 \*\B1 
     		  	- 29 \* \B0
     		  \right) \* \expF 
     		  + 17\* \, \expA 
     	\bigg]
\nn \\[1mm] & & \hspace{-2cm}\mbox{}
     	+  2\* \cfs \* \caf\* \bigg[
     		  \left(
     		  	  54 \*( \B2+ \B1)
     			+ (43-24 \* \z2) \* \B0
     		\right)\* \, \expF 
    			- 4\*\,( 31-6 \* \z2)\* \, \expA 
    		\bigg]
\nn \\[1mm] & & \hspace{-2cm}  \mbox{}   		
    		+ 40\*\,\b0\* \, \cafs \* \bigg[ 
    			  \B2\* \, \expF 
    			  - \expA 
    		\bigg]
     	+ 16\*\caft \* \bigg[
     		  (2+3 \* \z2)\*\B2\* \, \expF 
     		- (20-9 \* \z2) \* \expA
     	\bigg]
\nn \\[1mm] & & \hspace{-2cm}\mbox{}
     	- 16\*\cf \* \cafs \* \bigg[
     		  \left(
     		  	  (19-3 \* \z2) \* \B2 
     		  	- (2-3 \* \z2) \* \B0
     		  \right) \* \expF 
     		  + 4\* \,(7-3 \* \z2)\* \, \expA
     	\bigg]
	\Big\}
\nn \\[1mm] & & \hspace{-2.5cm}
	+ \:\frac{ \nf\*\ar^3 \*\b0 \* \ln^3\, \Ntil}{6\*\,\caf}\* 
	\Big\{
		- \, \b0\* \, \cf \* \left(
			  \B1
			- \B0
		\right) \* \expF 
		- 8 \*\cafs\*\, \expA
      	- 20\*\cf \* \caf \* \expA
\nn \\[1mm] & & \hspace{-2.5cm}
     	+ \:6\*\cfs \* \bigg[
     		  \left( 
     		  	  \B1
     		  	+ \B0
     		  \right) \* \expF 
     		  - 2\* \, \expA 
     	\bigg]
	\Big\}
	+ \frac{ \nf\*\ar^4\* \bb02\* \ln^5\, \Ntil}{9\*\caf }\* 
	\bigg[ 
		  \cfs \* \B0 \* \expF 
		- \cas\* \expA
	\bigg]
\label{eq:c2gNL}
\eea
and 
\bea
N\:\!C_{\phi,\rm q}^{}(N,\as) &\!=\!& 
	- \frac{\cf}{2\*\,\caf\*\ln \Ntil}\*\left[
		  \expA\*\Bb0
		- \expF
	\right]
	- \: \frac{\cf\*\left(\b0-3\*\cf\right)}{8\*\,\cafs\*\ln^2\Ntil}\*\left[
		  \expA\*\Bb0
		- \expF
	\right]
\nn \\[1mm] & & \hspace{-2.5cm}
	- \: \frac{\cf\*\ar}{4\*\,\caf}\*\: 
	\Big\{
		  6\*\cf\*\bigg[
		  	  \expA\*\left(
		  	  	  \Bb1
		  	  	+ 2\art^{-1}\*\Bb{-1}
		  	  \right) 
		  	- \expF
		\bigg]
	   		- \, 4\*\caf\*\bigg[
   			  2\*\expA\*\Bb1
   			- 3\*\expF
   		\bigg] 
\nn \\[0.5mm] & & \hspace{-2cm} \mbox{}
		- \,\b0\*\bigg[
			  \expA\*\left(
			  	  \Bb1
			  	- 2\*\Bb0
			  	+ 4\art^{-1}\*\Bb{-1}
			  	+ \art^{-1}\*\Bb{-2}
			\right) 
			+ \expF
		\bigg]
   	\Big\}
\nn \\[1mm] & & \hspace{-2.5cm}
	- \: \frac{\cf\*\ar^2\*\b0\*\ln^2\Ntil}{3\*\,\caf}\*
	\left[
		  \ca\*\expA\*\Bb0
		- \cf\*\expF
	\right]
	- \:\frac{\cf}{32\* \ln^3 \Ntil\*\caft}\*
	\Big\{ 
     	\bb02\*\left( 
     		  \frac{1}{3}\*\Bb{-3}
     		+ \frac{1}{8}\*\Bb{-4}
     	\right) \* \expA
\nn \\ & & \hspace{-2cm}
     	+ \b0 \*(\b0-3 \* \cf)\*
     	\left( 
     		  \Bb{-2}
     		+ \Bb{-3}
     	\right)\* \expA
 		+ (\b0-3 \* \cf)^2\*\bigg[
		  	  \left( 
		  	  	  \Bb0
		  	  	- \Bb{-1}
		  	  	+ 2 \* \Bb{-2}
		  	\right)\* \expA 
     		- \expF 
     	\bigg] 
     \Big\}
 \nn \\[1mm] & & \hspace{-2.5cm}
	+ \:\frac{\cf\*\ar}{96\*\,\cafs\*\ln \Ntil\:}\* 
	\Big\{
  		  3\*\b0^2\* \bigg[
  		  	  \left(
  		  	  	  20 \* \Bb1
  		  	  	- 16 \* \Bb0
  		  	  	+ 7 \* \Bb{-1}
  		  	  	+ 2 \* \Bb{-2}
  		  	  \right) \* \expA 
   			- 2\* \, \expF 
   		\bigg]
 \nn \\ & & \hspace{-2cm}\mbox{}
	     + 54\*\cfs\* \bigg[
	     	  \left(
	     	  	  10\* \Bb1
	     	  	- 7 \* \Bb0
	     	  \right) \* \expA 
	     	- 2\* \, \expF
	     \bigg]   		
    		- 18\* \b0 \* \cf \* \bigg[
    			  \left(
    			  	  22\* \Bb1
    			  	- 16 \* \Bb0
    			  	+ 3 \* \Bb{-1}
    			  \right) \* \expA 		
		  \nn \\[1mm] & & \hspace{-2cm} \mbox{}
 		   - 4\* \, \expF 
 		\bigg]
     	- 32\*\b0\* \,\caf \* \bigg[
     		  \left(
     		  	  2 \* \Bb1
     		  	+ 3 \* \Bb0
     		  	+ 2 \* \Bb{-1}
     		  \right) \* \expA 
     		- 5\* \, \expF  
     	\bigg]
\nn \\[1mm] & & \hspace{-2cm}\mbox{}
     	+ 16\*  \cf\*\caf \* \bigg[ 
     		  \left(
     		  	  16 \* \Bb1
     		  	- 3 \* (1-2 \* \z2) \* \Bb0
     		  	- (2-3 \* \z2) \* \Bb{-1}
     		  \right) \* \expA 
     		- (13+ 6 \* \z2) \* \expF
     	\bigg]
\nn \\[1mm] & & \hspace{-2cm}\mbox{}
     	- 16\*\cafs\* \bigg[ 
     		  \left(
     		  	  4 \*(5+3 \* \z2) \* \Bb1
     		  	- 6 \*(1+2 \* \z2) \* \Bb0
		      	+ (2-3 \* \z2) \* \Bb{-1}
		      \right) \* \expA 
		    - 14\* \, \expF
		\bigg]
	\Big\}
\nn \\[1mm] & & \hspace{-2.5cm}
	+ \:\frac{ \cf\*\ar^2\* \ln \Ntil}{24\*\,\cafs} \* 
	\Big\{
     	- 2\* \b0\*\, \cf\* \caf \* \bigg[
     		  \left(
     		  	- 9 \* \Bb2
     		  	+ 18 \* \Bb1
     		  	+ 17 \* \Bb0
     		  	+ 6 \* \Bb{-1}
     		  \right) \* \expA 
    			- 38\* \, \expF 
    		\bigg]
\nn \\ & & \hspace{-2cm}\mbox{}
   	   +   6\*\b0 \*\cfs \* \bigg[
   	    		  \left( 
   	    		  	  \Bb0
   	    		  	- 2 \* \Bb{-1}
   	    		  \right) \* \expA 
   	    		- \expF
   	    	\bigg]
		- \bb02 \* \cf\* \bigg[
			  \left(
			  	  2 \* \Bb0
			  	- 4 \* \Bb{-1}
			  	- \Bb{-2}
			  \right) \* \expA 
			- 2 \* \expF
		\bigg]
\nn \\[1mm] & & \hspace{-2cm}\mbox{}
     	- \bb02 \*\caf \* \bigg[
     		  \left(
     		  	  \Bb2
     		  	- 6 \* \Bb1
     		  	+ 14\*\Bb0
     		  	- 4 \* \Bb{-1}
     		  \right.
   			  \left.
   			  	- \Bb{-2}
   			  \right) \* \expA 
   			- 5\* \, \expF 
   		\bigg]
\nn \\[1mm] & & \hspace{-2cm}\mbox{}
      	- 2\*\cfs\* \caf\* \bigg[
      		  \left(
      		  	  54 \* \Bb2
      		  	+ 8 \*(2-3 \* \z2) \* \Bb0
      		  \right) \* \expA 
      		- (43-24 \* \z2)\* \, \expF 
      	\bigg]
\nn \\[1mm] & & \hspace{-2cm}\mbox{}
      	+ 8\*\caft\* \bigg[
      		  \left(
      		  	  2 \*(2-9 \* \z2) \* \Bb2
      		  	- 2 \*(2-3 \* \z2) \* \Bb0
      		  \right) \* \expA 
    			+ 12\*(1+2 \* \z2)\* \, \expF 
    		\bigg]
\nn \\[1mm] & & \hspace{-2cm}\mbox{}
      	- 8\*\cf \* \cafs \* \bigg[
      		  \left(
      		  	  (20-6 \* \z2) \* \Bb2
    			  	+ 4 \*(2-3 \*\z2) \* \Bb0
    			  \right) \* \expA 
    			+ (23+6 \* \z2)\* \, \expF 
    		\bigg]
\nn \\[1mm] & & \hspace{-2cm}\mbox{}
      	+ 8\* \b0 \* \cafs\* \bigg[
      		  \left(
      		  	  2 \* \Bb2
      		  	+ 6 \* \Bb1
      		  	- 5 \* \Bb0
      		  \right) \* \expA 
      		- 9\* \, \expF 
      	\bigg]
	\Big\}
\nn \\[1mm] & & \hspace{-2.5cm}
  	+ \:\frac{\cf\*\ar^3\*\b0 \* \ln^3 \Ntil}{6\*\,\caf}\*  
  	\Big\{ 
   		  \cf\*(\b0-6 \* \cf)\*\bigg[ 
   		  	  \Bb1 \* \expA 
   		  	- \expF
   		\bigg]
      	- 3\* \b0\*\cf\*\bigg[
      		  \Bb0 \* \expA 
      		- \expF 
      	\bigg]
\nn \\ & & \hspace{-2cm}\mbox{}
     	+ \b0\*\caf \*\left( 
     		  \Bb1 
     		- 3 \* \Bb0
     	\right) \* \expA
    		+ 8 \* \cafs \* \Bb1 \* \expA
    		+ 2 \* \cf\*\caf \*\bigg[ 
    			  \Bb1 \* \expA 
    			- 6 \* \expF
    		\bigg]
	\Big\}
\nn \\[1mm] & & \hspace{-2.5cm}
  	- \:\frac{ \cf\*\ar^4\* \bb02\* \ln^5 \Ntil}{9\*\,\caf} \* 
  	\bigg[ 
  		  \cas \* \Bb0 \* \expA
  		- \cfs \* \expF
  	\bigg]
\label{eq:cphiqNL}
\eea
with ${\cal E}_{A,F} \,=\,\exp(2\ar C_{A,F} \ln^2\Ntil)$. 
The first term in these results represent the LL contributions, the second 
to forth terms the NLL corrections, and the rest the new NNLL expressions.

%
\setcounter{equation}{0}
\section{NNLL resummation in semi-inclusive $e^+e^-$ annihilation}
\label{sec:SIA}
We now address the final-state (`time-like') off-diagonal splitting functions, 
which have not been presented at NNLL accuracy before, for the NLL expressions 
see Ref.~\cite{LoPresti:2012rg}. 
Since these quantities are closely related to their initial-state 
(`space-like') counterparts, the formulae for the two cases are very similar. 
Hence it is convenient to express the time-like results via their difference 
with respect to the corresponding space-like results as presented in the 
previous section as
\bea
\label{eq:diffpqgTSNNL}
       \frac{N}{\nf}\,P^{\,\rm T}_{\rm qg}(N,\as) 
 \,-\, \frac{N}{\cf}\,P^{\,\rm S}_{\rm gq}(N,\as) &\!=\!&  
	  8\,\*\ar^2\*\,\caf\* \ln \Ntil\*\:\B1
\nn \\[-0.5mm] & & \hspace{-3.5cm}
	+ \,\ar^2\* 
    \Big\{
		  4\,\*\ar\* \left(
		  	  6\,\*\cf
		  	- \b0
		\right)\*\caf\*\,\ln^2\Ntil\*\: \B2
  		- 2\* \left(
  			  24\,\*\caf\*\z2
  			+ 3\,\*\cf
  			- \b0
  		\right)\*\,\B1
\nn \\[-1mm] & & \hspace{-2.5cm} 
		\,+ \,6\* \left(
			  4\,\*\caf\*\z2 
			+ 2\,\*\cf-\b0
		\right)\*\,\B0
  		-\,\: \b0\*\,\B{-1}
 	\Big\}
\; , \\[2mm]
\label{eq:diffpgqTSNNL}
       \frac{N}{\cf}\:\!P^{\,\rm T}_{\rm gq}(N,\as) 
 \,-\, \frac{N}{\nf}\:\!P^{\,\rm S}_{\rm qg}(N,\as) &\!=\!&  
 	  8\,\*\ar^2\*\,\caf\* \ln \Ntil\*\:\Bb1
\nn \\[-0.5mm] & & \hspace{-3.5cm}
	+ \,\ar^2\*  
    \Big\{
		  4\,\*\ar\* \left(
		  	  6\,\*\cf
		  	- \b0
		\right)\*\caf\*\,\ln^2\Ntil\*\: \Bb2
  		+ 2\* \left(
  			  24\,\*\caf\*\z2 
  			- 3\,\*\cf
  			+ \b0
  		\right)\*\,\Bb1
\nn \\[-1mm] & & \hspace{-2.5cm} 
		\,-\, 2\* \left(
			  12\,\*\caf\*\z2 
			- 6\,\*\cf 
			+ \b0
		\right)\*\,\Bb0
  		+\,\: \b0\*\,\Bb{-1} 
	\Big\}
\; . 
\eea
The difference between the LL terms of the time-like and space-like splitting 
functions is zero (after removing the overall leading-order colour factors) 
\cite{AV2010}, the difference of the NLL terms is given by the respective 
first lines of Eqs.~(\ref{eq:diffpqgTSNNL}) and (\ref{eq:diffpgqTSNNL}), and 
the remaining terms represent the difference between the NNLL contributions.
  
\begin{figure}[htb]
\vspace{1mm}
\centerline{\epsfig{file=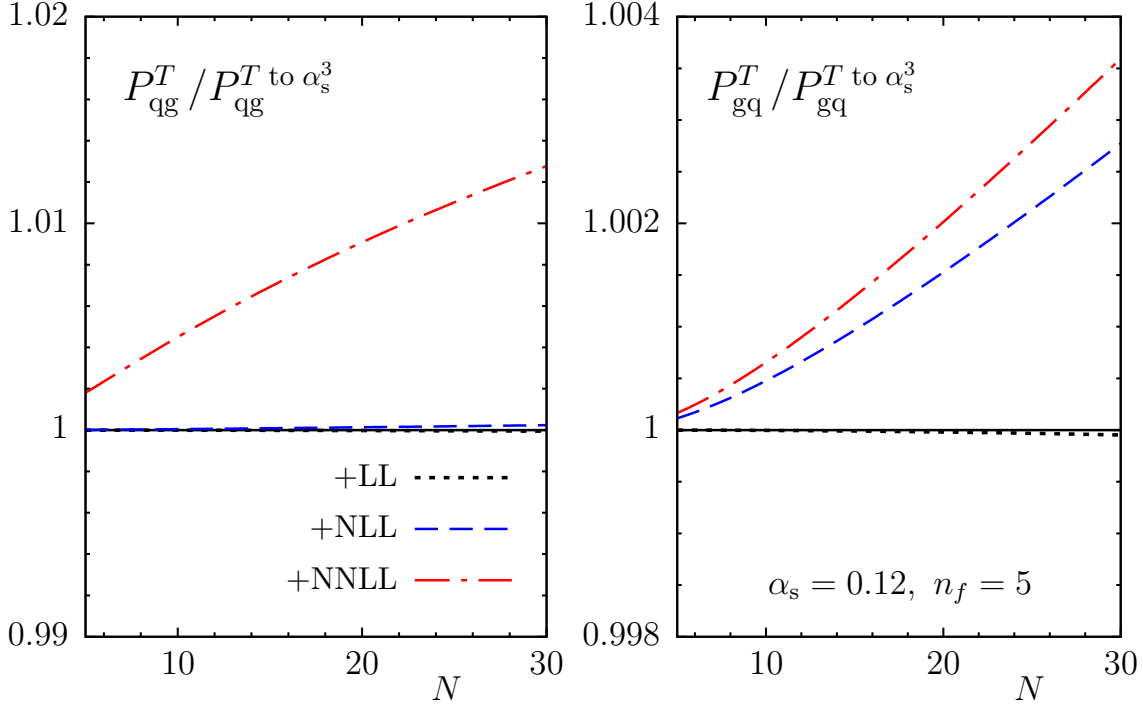,width=15cm,angle=0}}
\vspace{-2mm}
\caption{The relative size of the leading-logarithmic (LL), next-to-leading
 logarithmic (NLL) and next-to-next-to-leading logarithmic (NNLL) higher-order
 large-$N$ corrections  to the  NNLO off-diagonal splitting functions
 $P_{ij}^{\,T}(N)$ at a  typical high-scale  reference point.
\label{Figure1}
\vspace{0mm}
}
\end{figure}

\noindent
In Fig.~\ref{Figure1} the numerical size of the corrections beyond order
$\ar^{\,3}$ is illustrated at a scale $\Qs \simeq M_Z^{\,2}$, where these
corrections are entirely dominated by the $\ar^{\,4}$ terms. The presently
known contributions are small, but more large-$N$ terms are needed to
arrive at quantitatively reliable results.

These $\ar^{\,4}$ contributions to  Eqs.~(\ref{eq:diffpqgTSNNL}) and 
(\ref{eq:diffpgqTSNNL}), recall the normalization of $\ar$ in 
Eq.~(\ref{Pexp}), read
\bea
\label{eq:Pqg3DL0}
  {1\over\nf}\: P_{\rm qg}^{\,T\,(3)}(x) &\! = \!&
%
    \mbox{\hspn}
     \;\;\;\; \ln^{\,5}\*\! \x1 \* \Big[\,
            \frac{22}{27}\: \* \caft 
          \,- \, \frac{14}{27}\: \* \cafs \* \cf
          \,- \, \frac{4}{27}\: \* \cafs \* \nf
          \Big]
\nn \\[1mm] & & \mbox{\hspn}
     + \ln^{\,4}\*\! \x1 \* \Big[\,
            \Big( \, \frac{1432}{81}
            \,+\, \frac{64}{9}\: \* \z2\! \Big)\, \* \caft 
          \,+\, \Big( \, \frac{1471}{54}
          \,-\, 8 \* \z2\! \Big)\, \* \cafs \* \cf 
\nn \\[1mm] & & \mbox{}
          \,-\, \frac{49}{81}\: \* \caf \* \cfs 
          \,-\, \frac{16}{3}\: \* \cafs \* \nf
          \,+\, \frac{17}{81}\: \* \caf \* \cf \* \nf
          \,+\, \frac{32}{81}\: \* \caf \* \nfs
          \Big]
\nn \\[2mm] & & \mbox{\hspn}
      + {\cal O} \left( \ln^3 \! \x1 \right)
 \:\: ,
\\[4mm]
\label{eq:Pgq3DL0}
  {1\over\cf}\: P_{\rm gq}^{\,T\,\,(3)}(x) &\! = \!&
%
     \mbox{\hspn}
     \;\;\;\; \ln^{\,5}\*\! \x1 \* \Big [ \:
          \,-\, \frac{26}{27}\: \* \caft
          \,-\, \frac{14}{27}\: \* \cafs \* \cf
          \,-\, \frac{4}{27}\: \* \cafs \* \nf
          \Big]
\nn \\[1mm] & & \mbox{\hspn}
     + \ln^{\,4}\*\! \x1 \* \Big [ \,
            \Big( \, \frac{469}{27}
            \,-\, \frac{128}{9}\: \* \z2\! \Big)\, \* \caft 
          \,+\, \Big( \, \frac{5317}{162}
          \,-\, 8 \* \z2\! \Big)\, \* \cafs \* \cf
\nn \\[1mm] & & \mbox{}
          \,-\, \frac{13}{81}\: \* \caf \* \cfs 
          \,-\, \frac{212}{81}\: \* \cafs  \* \nf
          \,+\, \frac{17}{81}\: \* \caf \* \cf \* \nf
          \,-\, \frac{4}{81}\: \* \caf \* \nfs
          \Big]
\nn \\[2mm] & & \mbox{\hspn}
      + {\cal O} \left( \ln^3 \! \x1 \right)
\eea
after transformation to $x$-space using the second line of Eq.~(\ref{LTrf}).
The corresponding NLO and NNLO expressions can be found in Ref.~\cite{AMV1}.
Note that the coefficients of $\,\ln^{\,6}\x1\,$ are zero for both the
space-like and time-like functions. This is due to fact that the LL 
contributions at order $\ar^{\,n+1}$ are proportional to the Bernoulli 
numbers $B_n$ which vanishes for all odd values \mbox{$n>1$}.
Eqs.~(\ref{eq:Pqg3DL0}) and (\ref{eq:Pgq3DL0}) can be used, e.g., as a check 
for future fourth-order Feynman diagram calculations of the 
next-to-next-to-next-to-leading order (N$^{3}$LO) time-like splitting function
or, before, for use with other partial information on these quantities.  

Also for the SIA coefficient functions, the large-$N$ expressions can be 
conveniently presented via their differences with the corresponding 
quantities in DIS. 
In this form the results for transverse and $\phi$-exchange coefficient 
functions $C_{T,\rm g}$ and $C^{\,\rm T}_{\phi,\rm q}$ can be written in
a surprisingly compact form,
\bea
       \frac{N}{\cf}\:C_{T,\rm g}(N,\as)  
 \,-\, \frac{N}{\nf}\:C_{2,\rm g}(N,\as)  &\!=\!& 
	  2\* \,\ar\* \, \expF\* \, \B1
\nn \\ & & \hspace{-45mm}
	+ \,\frac{\ar}{4\*\, \caf\*\ln\, \Ntil}\:\* 
	\Big\{ 
    		  \bigg[ 
    		  	  \b0\*\left(
    		  	  	  4\*\B1
    		  	  	- 4\* \B0
    		  	  	+ \B{-1}
    		  	\right)
    		  	+ 12\*\,\cf\*\left(
    		  		- \B1
    		  		+ (1+2\*\z2)\* \B0
    		  	\right) 
\nn \\ & & \hspace{-35mm}
            + \,24\,\*\caf\*\z2\*\left(
            		  2\*\B1
            		- \B0
            	\right)
          \bigg]\*\expF 
        - 24\*\,\ca\*\,\z2\*\,\expA 
	\Big\}
\nn \\ & & \hspace{-45mm}
	- \,\ar^2\*\ln\,\Ntil\: \* 
	\Big\{  
		  \b0\*\B2
		- 6\*\,\cf\*\left(
			  \B2
			+ \B1
		\right) 
	\Big\}\*\expF
	+ \frac{4}{3}\,\* \ar^3\*\cf\*\b0\*\ln^3\, \Ntil \*\expF\*\B1
\label{eq:diffTS2T}
\; , \\[4mm]
       \frac{N}{\nf}\:C^{\,\rm T}_{\phi,\rm q}(N,\as)  
 \,-\, \frac{N}{\cf}\:C^{}_{\phi,\rm q}(N,\as) &\!=\!& 
	-  2\* \,\ar\* \, \expA\* \, \Bb1
\nn \\ & & \hspace{-45mm}
	+ \,\frac{\ar}{4\*\, \caf\*\ln\, \Ntil}\:\* 
	\Big\{ 
		\bigg[ 
			- \b0\*\left(
				  4\*\Bb1
				- 4\* \Bb0 
				- \Bb{-1}
			\right) 
        		+ 12\*\,\cf\*\left(
        			  \Bb1 
        			- (1+2\*\z2)\* \Bb0
        		\right) 
\nn \\ & & \hspace{-35mm}
            + 48\*\,\caf\*\z2\*\left(
            		  \Bb1
            		- \Bb0
            	\right)
        \bigg]\*\expA 
		+ 24\*\,\cf\*\,\z2\*\,\expF 
	\Big\}
\nn \\ & & \hspace{-45mm}
	+ \,\ar^2\*\ln\,\Ntil \* 
	\Big\{
		  \b0\*\left(
		  	  \Bb2
		  	- 2\*\Bb1
		\right) 
		- 6\*\,\cf\*\Bb2 
	\Big\}\*\expA
	- \frac{4}{3}\:\* \ar^3\*\ca\*\b0\*\ln^3\, \Ntil \*\expA\*\Bb1
\; . 
\label{eq:diffTSphi}
\eea
As for the splitting functions, the difference between the space- and 
time-like coefficient functions is zero at LL accuracy after dividing out 
the LO colour factors. The first line of Eq.~(\ref{eq:diffTS2T}) is the 
NLL difference and the remaining terms represent the difference of the 
NNLL corrections. 

The corresponding $x$-space expressions for the coefficient function 
$C_{T,\rm g}$ at order $\ar^{\,3}$ and $\ar^{\,4}$ read
\bea
\label{eq:cg3DL0}
 {1\over\cf}\: \* c_{T,{\rm g}}^{\,(3)}(x) &\! = \!&
%
    \mbox{\hspn}
     \;\;\;\; \ln^{\,5}\*\! \x1 \* \Big[\,
           \frac{2}{3}\: \*\cas
          \,+\, \frac{10}{3}\: \* \cfs
          \Big]
\nn \\[1mm] & & \mbox{\hspn}
     + \ln^{\,4}\*\! \x1 \* \Big[\,
            \frac{7}{27} \* \ca \* \nf
          \,-\, \frac{269}{54}\: \* \cas
          \,+\, \frac{17}{27}\: \* \cf \* \nf
          \,-\, \frac{338}{27}\: \* \cf \* \ca
          \,-\, \frac{97}{18}\: \* \cfs
          \Big]
\nn \\[1mm] & & \mbox{\hspn}
     + \ln^{\,3}\*\! \x1 \* \Big[\,
          \Big(\, \frac{2990}{81}
          \,-\, \frac{16}{9}\:  \* \z2\Big)\,\* \cas
          \,+\,\Big(\, \frac{3652}{81} 
          \,-\, \frac{88}{9}\:  \* \z2\Big)\,\* \cf \* \ca
\nn \\[1mm] & & \mbox{}
          \,-\, \Big(\,\frac{41}{9}\: 
          \,+\, \frac{112}{9}\:  \* \z2\Big)\,\* \cfs
          \,-\, \frac{140}{81}\: \* \ca \* \nf
          \,-\, \frac{436}{81}\: \* \cf \* \nf
          \Big]\;\;
\nn \\[2mm] & & \mbox{\hspn}
      + {\cal O} \left( \ln^2 \! \x1 \right)
\; , 
\eea
a result that has already been presented in Ref.~\cite{LoPresti:2012rg}, and
\bea
\label{eq:cg4DL0}
 {1\over\cf}\: \* c_{T,{\rm g}}^{\,(4)}(x) &\! = \!&
%
    \mbox{\hspn}
     \;\;\;\; \ln^{\,7}\*\! \x1 \* \Big[\,
           \frac{46}{135}\:  \* \cat
          \,+\, \frac{14}{45}\: \* \cf \* \cas
          \,-\, \frac{14}{45}\: \* \cfs \* \ca
          \,+\, \frac{314}{135}\: \* \cft
          \Big]
\nn \\[1mm] & & \mbox{\hspn}
     + \ln^{\,6}\*\! \x1 \* \Big[\,
           \frac{112}{405}  \* \cas \* \nf
          \,-\, \frac{1696}{405}\: \* \cat
          \,+\, \frac{106}{405}\: \* \cf \* \ca \* \nf
          \,-\, \frac{703}{162}\: \* \cf \* \cas
\nn \\[1mm] & & \mbox{}
          \,+\, \frac{502}{405}\: \* \cfs \* \nf
          \,-\, \frac{5407}{405}\: \* \cfs \* \ca
          \,-\, \frac{59}{10}\: \* \cfs
          \Big]
\nn \\[1mm] & & \mbox{\hspn}
     + \ln^{\,5}\*\! \x1 \* \Big[\,
        \Big( \,  \frac{75403}{1620} 
          \,-\, \frac{149}{15}\: \* \z2\Big) \* \cat \,
    \,+\, \Big( \, \frac{22937}{648} 
          \,+\, \frac{52}{15}\:  \* \z2\Big) \* \cf \* \cas\,
\nn \\[1mm] & & \mbox{}
    \,+\, \Big( \, \frac{10055}{108} 
          \,-\, \frac{99}{5}\:  \* \z2\Big)\* \cfs \* \ca \,
     \,-\,\Big( \, \frac{143}{120}\: 
          \,+\, \frac{326}{15}\: \* \z2\Big)\* \cft  \,
          \,+\, \frac{23}{405}\: \* \ca \* \nfs
\nn \\[1mm] & & \mbox{}
          \,-\, \frac{521}{135}\: \* \cas \* \nf
          \,+\, \frac{97}{405}\: \* \cf \* \nfs
          \,-\, \frac{3503}{540}\: \* \cf \* \ca \* \nf
          \,-\, \frac{6013}{540}\: \* \cfs \* \nf
          \Big]
\nn \\[2mm] & & \mbox{\hspn}
      + {\cal O} \left( \ln^4 \! \x1 \right)
\; .
\eea
The corresponding results for scalar-exchange SIA are given by
\bea
\label{eq:cq3DL0}
 {1\over\nf}\: \*  c_{\phi,{\rm q}}^{\,T\,\,(3)}(x) &\! = \!&
%
     \mbox{\hspn}
     \;\;\;\; \ln^{\,5}\*\! \x1 \* \Big [ \:
            \frac{10}{3}\: \* \cas
          \,+\, \frac{2}{3}\: \* \cfs
          \Big]
\nn \\[1mm] & & \mbox{\hspn}
     + \ln^{\,4}\*\! \x1 \* \Big [ \,
            \frac{47}{27}\: \* \ca \* \nf
          \,-\, \frac{517}{54}\: \* \cas
          \,+\, \frac{13}{27}\: \* \cf \* \nf
          \,-\, \frac{310}{27}\: \* \cf \* \ca
          \,-\, \frac{55}{6}\: \* \cfs
          \Big]
\nn \\[1mm] & & \mbox{\hspn}
     + \ln^{\,3}\*\! \x1 \* \Big [ \,
         \Big ( \, \frac{6554}{81} 
          \,-\, \frac{104}{9}\: \* \z2 \Big ) \,\* \cas 
          \,+\,\Big ( \, \frac{6139}{81} 
          \,-\, \frac{248}{9}\: \* \z2\Big ) \,\* \cf \* \ca
\nn \\[1mm] & & \mbox{}
          \,+\, \Big ( \, \frac{64}{3}
          \,+\, \frac{208}{9}\:  \* \z2\Big ) \,\* \cfs
          \,+\, \frac{16}{27}\: \* \nfs
          \,-\, \frac{1268}{81}\: \* \ca \* \nf
          \,-\, \frac{970}{81}\: \* \cf \* \nf
          \Big]
\nn \\[2mm] & & \mbox{\hspn}
      + {\cal O} \left( \ln^2 \! \x1 \right)\;\;,\\[5mm]
\label{eq:cq4DL0}
 {1\over\nf}\: \*  c_{\phi,{\rm q}}^{\,T\,\,(4)}(x) &\! = \!&
%
     \mbox{\hspn}
     \;\;\;\; \ln^{\,7}\*\! \x1 \* \Big [ \:
           \frac{314}{135}\: \* \cat 
          \,-\, \frac{14}{45}\: \* \cf \* \cas 
          \,+\, \frac{14}{45}\: \* \cfs \* \ca 
          \,+\, \frac{46}{135}\: \* \cft 
          \Big]
\nn \\[1mm] & & \mbox{\hspn}
     + \ln^{\,6}\*\! \x1 \* \Big [ \,
          \frac{1004}{405}\: \* \cas \* \nf
          \,-\, \frac{5522}{405}\: \* \cat 
          \,+\, \frac{2}{405}\: \* \cf \* \ca \* \nf
          \,-\, \frac{6403}{810}\: \* \cf \* \cas 
\nn \\[1mm] & & \mbox{}
          \,+\, \frac{254}{405}\: \* \cfs \* \nf
          \,-\, \frac{2171}{405}\: \* \cfs \* \ca 
          \,-\, \frac{559}{90}\: \* \cft 
          \Big]
\nn \\[1mm] & & \mbox{\hspn}
     + \ln^{\,5}\*\! \x1 \* \Big [ \,
        \Big( \, \frac{194611}{1620} 
          \,-\, \frac{1183}{45}\:  \* \z2\Big) \* \cat \,
    \,+\, \Big( \, \frac{16873}{216} 
          \,-\, \frac{796}{ 45}\:  \* \z2\Big) \* \cf \* \cas\,
\nn \\[1mm] & & \mbox{}
    \,+\, \Big( \, \frac{103781}{1620} 
          \,-\, \frac{499}{45}\: \* \z2\Big) \* \cfs \* \ca \*\,
    \,+\, \Big( \, \frac{9649}{360} 
          \,+\, \frac{226}{15}\: \* \z2\Big) \* \cft \,
          \,+\, \frac{187}{135}\: \* \ca \* \nfs
\nn \\[1mm] & & \mbox{}
          \,-\, \frac{10846}{405}\: \* \cas \* \nf
          \,+\, \frac{53}{135}\: \* \cf \* \nfs
          \,-\, \frac{709}{60}\: \* \cf \* \ca \* \nf
          \,-\, \frac{20993}{1620}\: \* \cfs \* \nf
          \Big]
\nn \\[2mm] & & \mbox{\hspn}
      + {\cal O} \left( \ln^4 \! \x1 \right)
 \:\: .
\eea


As for $C_{\,2,\,q}$, $C_{\,T,\,q}$ and $C_{\phi,g}^{\,(T)}$, the leading 
(and subleading) $1/N^{\,k}$ parts of the DIS and SIA quark coefficient 
functions for $F_L$ are given by `non-singlet' contributions that have been 
derived and discussed in Refs.~\cite{MV3,MV5}. 
The gluon coefficient functions have an analytical structure analogous 
to Eqs.~(\ref{eq:diffpqgTSNNL}) and (\ref{eq:diffTS2T}), with 
the SIA (time-like, T) and DIS large-$N$ expressions for $C_{\,L,g}$ 
differing only at NNLL accuracy. The same holds for $C_{\,L,q\,}$,
see Eq.~(6.16) of Ref.~\cite{MV5}. We find
\bea
       \frac{N^{\,2}}{\cf}\:\!C^{\,\rm T}_{L,\rm g}(N,\as) 
 \,-\, \frac{N^{\,2}}{2\nf}\:\!C^{}_{L,\rm g}(N,\as) &\!=\!& 
    4\*\,\ar^2\*\,\cf\*\, \expF\* \B1\;
	+\; 48\*\,\ar^2\*\ca\*\z2\* \expA
\; ,
\label{eq:diffTSlong}
\eea
where the analytic NNLL expression for $C^{}_{L,\rm g}$ has already
been given in Eq.~(6.3) of Ref.~\cite{Almasy:2010wn}. 

The resulting third- and fourth-order NNLL threshold expansion of 
$C_{\,L,g}^{\,T}$ in $x$-space is given by
\bea
\label{eq:cL3DL2}
  \x1^{-1}\: c_{L,{\rm g}}^{\,T\,(3)}(x) &\! =\! & \:
      8\,\* \cf \* \cas\, \* \ln^{\,4}\*\! \x1
\nn \\[1mm] & & \mbox{\hspn}
      + \ln^{\,3}\*\! \x1\: \* \cf \* \Big[
              \,  \frac{20}{3}\, \* \cfs
           \,+\, \frac{52}{3}\, \* \cf \* \ca
           \,-\, \frac{952}{9}\, \* \cas
           \,+\, \frac{16}{9}\, \* \ca \* \nf
           \Big]
\nn \\[1mm] & & \mbox{\hspn}
       + \ln^{\,2}\*\! \x1\; \* \cf \* \Big[
           \left(62
           - 32\*\*\z2\right)\*\cfs
           - \left(\frac{784}{3}
           - 32\*\z2\right)\*\ca\*\cf
           + \frac{5720}{9}\*\cas
\nn \\ & & \mbox{}
            - \frac{224}{9}\*\ca\*\nf
            + \frac{16}{3}\*\cf\*\nf
  -64\*\,\nfs\*\:\dabcna\*\:\flg11\*\,\*\,(11+2\,\*\z2-12\*\,\z3)
   \Big]
\nn \\[2mm] & & \mbox{\hspn}
       + {\cal O} \left( \ln \! \x1 \right)
%
\; , \\[4mm] 
\label{eq:cL4DL2}
 \x1^{-1}\: c_{L,{\rm g}}^{\,T\,(4)}(x) &\! =\! & \:
     \frac{16}{3}\,\* \cf \* \cat\, \* \ln^{\,6}\*\! \x1 
\nn \\[1mm] & & \mbox{\hspn}
     + \ln^{\,5}\*\! \x1\: \* \cf \* \Big [ 
               \frac{20}{3}\, \* \cft
          \,+\, \frac{52}{3}\, \* \cf \* \cas 
          \,-\, \frac{1040}{9}\, \* \cat
          \,+\, \frac{32}{9}\, \* \cas \* \nf 
          \Big] 
\nn \\[1mm] & & \mbox{\hspn}
      + \ln^{\,4}\*\! \x1\; \* \cf \* \Big[
           \Big(\frac{323}{9} - \frac{160}{3}\,\*\z2\Big)\,\*\cft
          + \Big(\frac{536}{27}+ 16\,\*\z2\Big)\,\*\ca\* \cfs
\nn \\[1mm] & & \mbox{}
          - \Big(\frac{12629}{27} - \frac{160}{3}\,\*\z2\Big)\,\*\cf\*\cas
          + \Big(\frac{35380}{27} - 80\,\*\z2\Big)\,\*\cat
\nn \\[1mm] & & \mbox{}
          + \frac{154}{27}\,\*\cfs\*\nf
          + \frac{278}{27}\,\*\cf\*\ca\*\nf
          - \frac{2096}{27}\,\*\cas\*\nf
          + \frac{16}{27}\,\*\ca\*\nfs
\nn \\[1mm] & & \mbox{}
 -128\*\,\nfs\*\:\dabcna\*\:\flg11\*\,\ca\*\,(11+2\,\*\z2-12\*\,\z3)
  \Big]
\nn \\[1mm] & & \mbox{\hspn}
      + {\cal O} \left( \ln^3 \! \x1 \right)
\; .
\eea
Eq.~(\ref{eq:cL4DL2}) was already given in Ref.~\cite{LoPresti:2012rg} if, 
for brevity, without the $\flg11$ contribution (see Fig.~1 of Ref.~\cite{MVV6} 
for a typical DIS diagram contributing to this flavour structure) which, for 
photon-exchange SIA, corresponds to the charge factor
\beq
\label{flg11}
  fl_{11}^{\:\rm g}  \:=\: \langle e \rangle ^2 / \langle e^{\,2} \rangle 
  \quad \mbox{ with } \quad
  \langle e^{\,k} \rangle \:=\: \nf^{\!\!-1}\: \textstyle 
  \sum_{\,i=1}^{\,\nf}\: e_i^{\,k} \; ,
\eeq
where $e_i$ is the charge of the $i$-th effectively massless flavour in 
units of the proton charge. Analogous terms also contribute to $C_{T,g}$ 
at the same powers of $\ln\x1$, but there these logarithms are below NNLL
accuracy, recall Eqs.~(\ref{Cak-xto1}) and (\ref{CLk-xto1}). 
Note that different normalizations are used in the literature for the QCD 
group factor $d^{\:\!abc}d_{abc}/n_a= 3/8\; d^{\:\!abc}d_{abc}/n_c= 5/48.$ 
Also note that our normalization of both functions (as that of other 
recent articles) differs by a factor of $\frac{1}{2}$ from that of 
Refs.~\cite{Rijken:1996vr,Rijken:1996ns,MM06}, i.e., here the first-order 
large-$x$ limits read
\bea
  C_{T,g}(x,\ar) \,=\, 2\,\cf \ar \, \ln \x1 \,+\: \ldots 
  \quad \mbox{and} \quad
  C_{L,g}^{\,T}(x,\ar) \,=\, 4\,\cf \ar \, (1-x) \,+\: \ldots\;\;.
\label{eq:CTLlargexLO}
\eea

\begin{figure}[bh]
\vspace*{4mm}
\centerline{\epsfig{file=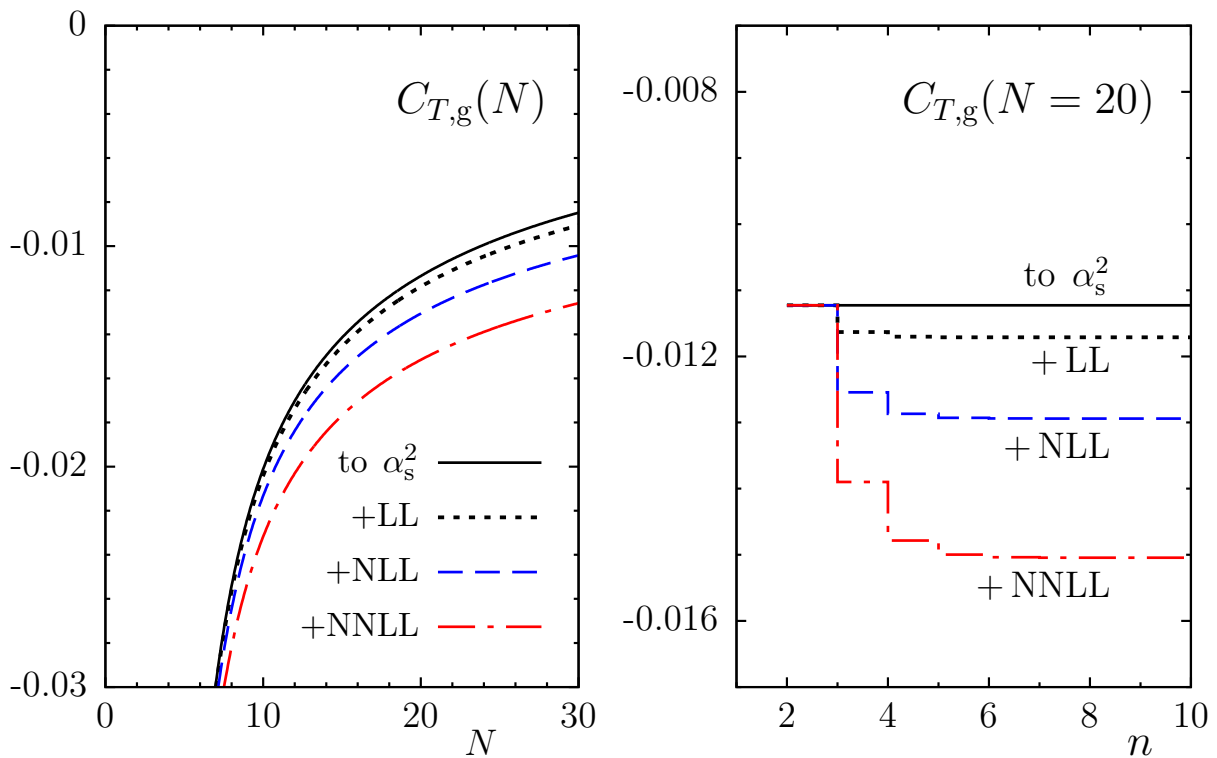,width=16cm,angle=0}}
\vspace*{-1mm}
\caption{Left: the large-$N$ behaviour of the second-order and resummed 
 \MSb-scheme gluon coefficient function for the transverse fragmentation 
 function at a high-scale reference point $\as \:=\: 0.12$ for $\nf = 5$ 
 light quark flavours.
 Right: the LL, NLL and NNLL contributions of the third to the tenth orders in 
 $\as$, added at the corresponding values of the abscissa, to those results at 
 $N=20$.\label{fig:CTg}}
\vspace*{1mm}
\end{figure}

\begin{figure}[tb]
\centerline{\epsfig{file=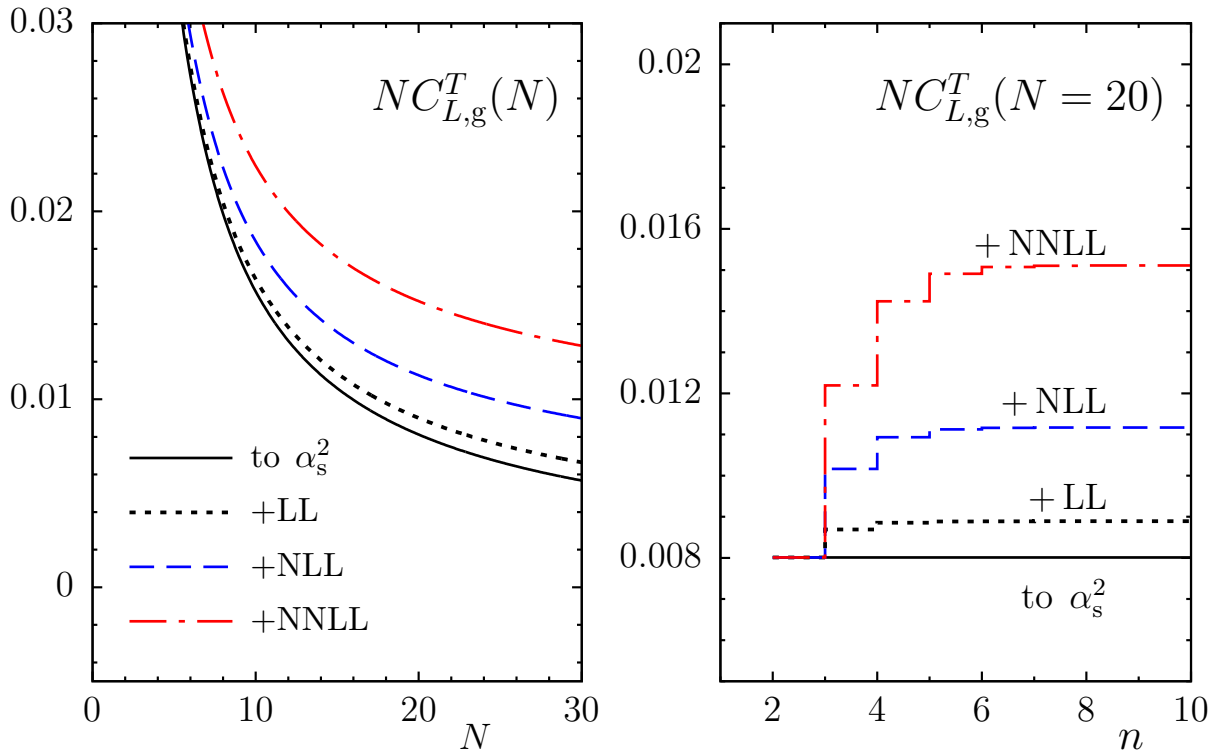,width=16cm,angle=0}}
\vspace*{-1mm}
\caption{As Fig.~\ref{fig:CTg}, but for the gluon coefficient function 
 of the longitudinal fragmentation function in photon-exchange SIA.
 All curves have been multiplied by $N$ for display purposes.
 \label{fig:CTLg}}
\vspace*{8mm}
\end{figure}

The results for $C_{T,g}$ in Eq.~(\ref{eq:diffTS2T}) and for $C_{L,g}^{\,T}$ 
in Eq.~(\ref{eq:diffTSlong}) are illustrated at the standard high-scale
reference point $\Qs \simeq M_Z^{\,2}$ (recall that we identify the
renormalization and factorization scales with $\Qs$ throughout this article)
in Fig.~\ref{fig:CTg} and Fig.~\ref{fig:CTLg}, respectively.
As for the splitting function in Fig.~\ref{Figure1}, the LL terms have a
small numerical effect. However, the overall (relative) size of the
corrections -- note that the effect of these coefficient functions is much
smaller than that of their quark counterparts \cite{MV3,MV5} -- is large here,
and contributions beyond order $\ar^{\,4}$ are not negligible as shown for
one moment $N$ in the right panels. 
Clearly higher terms in the large-$N$ expansion, or other information
complementing our results, are required to quantitatively establish the
size of the higher-order large-$N$ contributions to $C_{T,g}$ and 
$C_{L,g}^{\,T}$.
The results for $C_{\phi,q}^{\,T}$ in Eq.~(\ref{eq:diffTSphi}) are similar 
but not shown here, since this quantity is of mainly theoretical importance.

%
\section{Summary and outlook}
\label{sec:summary2}
Considerable progress has been made in the past seven years on the resummation 
of large-$x$ (or, in Mellin space, large-$N$) threshold logarithms \cite
{AV2010,Almasy:2010wn,LoPresti:2012rg,Vogt:2012gb,Presti:2014lqa,%
MV3,MV5,SMVV,avLL2010,dFMMV,%
Laenen:2008ux,Laenen:2008gt,Gardi:2010rn,Laenen:2010uz,White:2014qia,%
Bonocore:2014wua,Bonocore:2015esa,%
Grunberg:2009yi,Grunberg:2009vs,Grunberg:2011gx}
beyond those addressed by the soft-gluon exponentiation (SGE) \cite
{Sterman:1986aj,Catani:1989ne,Magnea:1990qg,Catani:1996yz,Contopanagos:1997nh}.
This holds for sub-leading contributions, in terms of powers of $\x1$ or $1/N$
for $x \ra 1$ or $N \ra \infty$, to quantities to which the SGE is applicable
for the leading terms, as well as for which the SGE is not applicable at all.

So far most of the explicit large-$x$ results for higher-order splitting 
functions and coefficient functions have been obtained by studying physical 
evolution kernels
\cite{MV3,MV5,SMVV,avLL2010,dFMMV,Grunberg:2009yi,Grunberg:2009vs,%
Grunberg:2011gx}
and the structure of unfactorized cross sections in dimensional regularization
\cite{AV2010,Almasy:2010wn,LoPresti:2012rg,Vogt:2012gb,Presti:2014lqa}
(see Refs.~\cite{AV2011,KVY} for an analogous small-$x$ resummation in SIA). 
The former approach is particularly suited for non-singlet quantities, i.e.,
quantities that only involve the quark-quark and gluon-gluon splitting 
functions which are stable in the threshold limit \cite{Korchemsky:1989si,%
DMS05,Moch:2004pa,Vogt:2004mw,Mitov:2006ic,Moch:2007tx}
in the standard \MSb\ factorization scheme adopted throughout this article. 
Where applicable beyond the leading logarithms, presently for inclusive 
deep-inelastic scattering (DIS) and semi-inclusive $e^+e^-$ annihilation (SIA),
the resummation via unfactorized cross sections leads to the same results
but is slightly more powerful even in those non-singlet cases.
The resummation of the off-diagonal quark-gluon and gluon-quark splitting
function is possible only in this second approach.

Both methods are, so far, less effective for lepton-pair and Higgs-boson
production, the former due to the lack of suitable flavour-singlet physical 
evolution kernels, the latter due to the different phase-space structure of 
these hadron collider processes. However, the complete analogy of the 
non-singlet physical kernels for SIA, DIS, the Drell-Yan process and Higgs 
production \cite{MV5,dFMMV} suggests that there is presently unearthed
resummation information also in the unfactorized expressions for the latter
two processes.

In the present article, we have first reconsidered the NNLL threshold
resummation of the flavour-singlet structure functions in DIS. This
resummation was performed already in Ref.~\cite{Almasy:2010wn} where,
however, a closed NNLL expression was given only for the simplest case,
the longitudinal structure function $\FL$. 
Here we have presented corresponding closed forms also for the off-diagonal
splitting functions $P_{qg}$ and $P_{gq}$ for the evolution of the parton
distributions of hadrons and for the corresponding coefficient functions for
$\Ftwo$ and $F_\phi$, where the latter occurs in DIS via Higgs-boson exchange 
in the heavy-top limit.

We have then extended this resummation to the theoretically closely related,
see Ref.~\cite{Mitov:2006ic,Moch:2007tx,AMV1}, case of SIA and presented the
NNLL resummation of the `time-like' splitting functions $P_{qg}^{\,T}$ and 
$P_{gq}^{\,T}$ for the evolution of final-state fragmentation distributions
and the corresponding coefficient functions for flavour-singlet fragmentation
functions. 
All these results can be expressed in terms of the Bernoulli functions
introduced in Ref.~\cite{AV2010} together with, for the coefficient functions, 
the quark and gluon leading-logarithmic soft-gluon exponentials. As already in 
Ref.~\cite{AV2010,Almasy:2010wn}, the analysis of physical kernels was useful
for finding the rather complicated closed NNLL expressions for the coefficient
functions.

We find that the resummation of the highest three logarithms, which leads
to small corrections for the off-diagonal splitting functions but large
effects for the corresponding coefficient functions, does not lead to 
numerically reliable results: the LL terms are always small, and the NLL
contributions mostly smaller than the NNLL `corrections'; a situation that, 
in fact, is similar to that in the SGE of the corresponding diagonal 
quantities Refs.~\cite{Moch:2005ba,MV4}. Yet we expect our results to become 
phenomenologically useful in connection with other efforts such as, e.g.,
the extension of the fourth-order calculations of Ref.~\cite{BC2006,%
VelizNS2,VelizhNS4,BCK2point} to higher moments and other quantities. 
Furthermore one may hope that the wealth of results derived here and 
before can provide some assistance to extending the reach of the non-SGE 
threshold resummation to higher logarithmic accuracies.

\subsection*{Acknowledgements}
This research has been supported by the German Research Foundation (DFG) 
through Sonderforschungsbereich Transregio 9, Computergest\"utzte 
Theoretische Teilchenphysik,
the Schweizer Nationalfonds (SNF) under grant 200020-162487
and the UK Science \& Technology Facilities Council (STFC) under grant 
numbers ST/J000493/1 and ST/L000431/1.
Our calculations were performed using the symbolic manipulation system 
FORM \cite{FORM3,TFORM,FORM4}.

%
{\footnotesize
\setlength{\baselineskip}{0.48cm}

}

\end{document}